\documentstyle[12pt,epsf,epsfig,rotating]{article}

\def\simlt{\stackrel{<}{{}_\sim}}

\def\beq{\begin{equation}}
\def\eeq{\end{equation}}
\def\bea{\begin{eqnarray}}
\def\eea{\end{eqnarray}}

\begin{document}

\newcommand{\sheptitle}
{Leptogenesis with Single Right-Handed Neutrino Dominance}

\newcommand{\shepauthor}
{M. Hirsch and S. F. King }

\newcommand{\shepaddress}
{Department of Physics and Astronomy,
University of Southampton, Southampton, SO17 1BJ, U.K.}

\newcommand{\shepabstract}
{We make an analytic and numerical study of leptogenesis in the framework 
of the (Supersymmetric) Standard Model plus the see-saw mechanism with a 
$U(1)$ family symmetry and single right-handed neutrino dominance. In 
presenting our analytic and numerical results we make a clear distinction 
between the theoretically clean asymmetry parameter $\epsilon_1$ and the 
baryon asymmetry $Y_B$. In calculating $Y_B$ we propose and use a fit to 
the solutions to the Boltzmann equations which gives substantially more 
reliable results than parametrisations previously used in the literature. 
Our results show that there is a decoupling between the low energy neutrino 
observables and the leptogenesis predictions, but that nevertheless 
leptogenesis is capable of resolving ambiguities within classes of models 
which would otherwise lead to similar neutrino observables. For example we 
show that models where the dominant right-handed neutrino is the heaviest 
are preferred to models where it is the lightest and study an explicit 
example of a unified model of this type. }

\begin{titlepage}
\begin{flushright}
hep-ph/0107014\\
\end{flushright}
\begin{center}
{\large{\bf \sheptitle}}
\bigskip \\ \shepauthor \\ \mbox{} \\ {\it \shepaddress} \\ \vspace{.5in}
{\bf Abstract} \bigskip \end{center} \setcounter{page}{0}
\shepabstract
\end{titlepage}

\section{Introduction}

Leptogenesis is an interesting mechanism which has been proposed to generate 
the observed baryon asymmetry of the Universe (BAU) \cite{yanagida1,luty}. 
The mechanism involves the out-of-equilibrium decay of a heavy right 
handed neutrino $N_R$. The net lepton number L produced in the decay is 
then reprocessed into baryon number B by anomalous (B+L) violating 
sphaleron interactions, which otherwise conserve (B-L) \cite{kuzmin}. 

The advantage of this mechanism is that the same physics that allows the 
right handed neutrinos to decay into light leptons is also responsible 
for a see-saw neutrino mass matrix \cite{seesaw}. 
This point of view has been 
strengthened by the latest experimental data on the solar neutrino 
problem by SNO \cite{SNO} and Super-Kamiokande 
\cite{superK} which, when combined, 
now seems to confirm the existence of a solar 
neutrino mass scale, and suggests active neutrino oscillations
based on either the LMA or the LOW solution \cite{bah2001}.
This in turn gives impetus to the see-saw mechanism.
Combining the see-saw mechanism with the experimental 
data \cite{SNO,superK} seems to favour scales for right handed 
neutrino masses $M_R$ in the range $10^{7}$ - $10^{16}$ GeV. 
There have been many studies of leptogenesis, all based on different
models, for example left-right symmetry, SO(10), and so on 
\cite{lepto}.

In this paper we study leptogenesis in the framework
of the (Supersymmetric) Standard Model plus the see-saw mechanism with
single right-handed neutrino dominance \cite{SRHND},
\cite{SRHND2}. SRHND is useful for both the LMA and the LOW
solution \cite{SRHND2} since it
leads to a natural neutrino mass hierarchy
in the presence of large mixing angles, and gives results
which are stable under radiative corrections
\cite{SRHNDRGE}. This provides
a relatively model independent approach which applies to a
large class of models with a natural hierarchy of neutrino masses
\cite{SelModels}.
Indeed in the case of the LOW solution, SRHND is almost inevitable
in order to maintain the large neutrino mass hierarchy
present in this case.

Within the SRHND framework we generalise previously presented 
analytic estimates for the mixing angles to the complex domain, 
and present new analytic results for the leptogenesis asymmetry parameter 
$\epsilon_1$ and discuss the insights which this leads to. We then 
introduce a $U(1)$ family symmetry \cite{flavour} and discuss our 
numerical approach to models of this kind. Our analytic results above 
are supported by the detailed numerical analysis of various texture 
models. Texture models involve unknown coefficients multiplying 
the expansion parameters, which implies some level of uncertainty 
in the predictions. In order to quantify this we perform a numerical 
scan over the unknown coefficients, to obtain distributions for 
predictions of neutrino masses, mixing angles as well as the 
predictions for $\epsilon_1$ and the baryon asymmetry $Y_B$, 
for different classes of models. 
In presenting our analytic and numerical results
we make a clear distinction between the theoretically clean
asymmetry parameter $\epsilon_1$ and the baryon asymmetry $Y_B$.
In calculating $Y_B$ we propose and use a fit to the solutions
to the Boltzmann equations which gives substantially more reliable results
than parametrisations previously used in the literature.
Using the numerical approach, 
supported by the analytic estimates, we then discuss two important 
aspects of leptogenesis, namely leptogenesis decoupling and 
leptogenesis discrimination.

We demonstrate explicitly that there is a {\em decoupling} between 
leptogenesis and the experimentally measurable neutrino parameters. 
Although such a result may be inferred by comparing the results from 
different individual models which have been proposed in the literature, 
the present paper represents the first attempt to systematically 
demonstrate this within a framework (SRHND) which can be plausibly 
applied to many different models. To support the decoupling claim we 
present examples of classes of models which give the same measurable 
neutrino parameters but have very different values for the CP 
asymmetry $\epsilon_1$. Leptogenesis decoupling implies that 
there is no relation for example between the size of the solar 
neutrino angle or MNS phase and the baryon asymmetry
predicted by leptogenesis.

On the other hand we show that leptogenesis is capable of 
{\em discriminating} between different models and thereby 
resolving ambiguities within classes of models giving the 
same low energy predictions. For example leptogenesis may 
resolve the ambiguity as to whether the dominant right-handed 
neutrino (the one chiefly responsible for the atmospheric 
neutrino mass in hierarchical models) is the heaviest or the 
lightest of the right-handed neutrinos. We show that within 
a standard hot big bang universe the models where the dominant 
right-handed neutrino is the heaviest are preferred and are more
consistent with the gravitino constraint on the reheat 
temperature $T_R\simlt 10^9 {\rm GeV}$ \cite{gravitino}. 

In section 2 we introduce our conventions, especially the use of 
the diagonal charged lepton and right-handed neutrino basis, the 
see-saw mechanism and the MNS matrix in this basis, and the standard 
model leptogenesis formulae in this basis. In calculating
the baryon asymmetry $Y_B$ in section 2.3
we present and use a new fit formula based on a Boltzmann analysis.
In section 3 we give our 
analytic results based on SRHND for the MNS parameters and
leptogenesis, which give important insights into the numerical
results which follow.
In section 4 we discuss our numerical approach to $U(1)$ family symmetry 
models. Section 5 is a discussion of the decoupling feature of
leptogenesis based on the calculation of the asymmetry parameter
$\epsilon_1$. 
In section 6 we discuss the calculation
of $Y_B$ for the models where the dominant right-handed neutrino
is the lightest, and show that such models are not consistent
with a standard hot big bang scenario. In section 7 we 
then discuss models where the dominant right-handed neutrino is
the heaviest and show that such models
can lead to successful leptogenesis. 
Section 8 concludes the paper.

\section{Conventions}

\subsection{The Diagonal Charged Lepton and Right-Handed Neutrino
Basis}

To fix the notation we consider the Yukawa terms with two Higgs doublets 
augmented by $3$ right-handed neutrinos, 
which, ignoring the quarks, are given by
\beq
{\cal L}_{yuk}=\epsilon_{ab}\left[\tilde{Y}^e_{ij}H_d^aL_i^bE^c_j
-\tilde{Y}^{\nu}_{ij}H_u^aL_i^bN^c_j + 
\frac{1}{2}\tilde{Y}_{RR}^{ij}\Sigma N^c_iN^c_j
\right] +H.c. 
\label{MSSM}
\eeq
where $\epsilon_{ab}=-\epsilon_{ba}$, $\epsilon_{12}=1$,
and the remaining notation is standard except that
the $3$ right-handed neutrinos $N_R^p$ have been replaced by their
CP conjugates $N^c_i$ 
and we have introduced a singlet field $\Sigma$ whose
vacuum expectation value (VEV) induces a heavy 
complex symmetric Majorana matrix 
$\tilde{M}_{RR}=<\Sigma>\tilde{Y}_{RR}$. 
When the two Higgs doublets get their 
VEVS $<H_u^2>=v_2$, $<H_d^1>=v_1$ we find the terms
\footnote{In the case of the standard model
we must replace 
one of the two Higgs doublets by the charge conjugate of the other,
$H_d=H_u^c$.}
\beq
{\cal L}_{yuk}= v_1\tilde{Y}^e_{ij}E_iE^c_j
+v_2\tilde{Y}^{\nu}_{ij}N_iN^c_j + 
\frac{1}{2}\tilde{M}_{RR}^{ij}N^c_iN^c_j +H.c.
\eeq
Replacing CP conjugate fields we can write in a matrix notation
\beq
{\cal L}_{yuk}=\bar{E}_Lv_1\tilde{Y^e}^\ast E_R
+\bar{N}_Lv_2\tilde{Y^\nu}^\ast N_R + 
\frac{1}{2}N^T_R\tilde{M}_{RR}^\ast N_R +H.c.
\eeq
It is convenient to work in the diagonal charged lepton basis
\beq
{\rm diag(m_e,m_{\mu},m_{\tau})}
= V_{eL}v_1\tilde{Y^e}^\ast V_{eR}^{\dag}
\eeq
and the diagonal right-handed neutrino basis
\beq
{\rm diag(M_{1},M_{2},M_{3})} =
V_{\nu_R}\tilde{M^\ast}_{RR}V_{\nu_R}^{T}
\eeq
where $V_{eL},V_{eR},V_{\nu_R}$ are unitary transformations.
In this basis the neutrino Yukawa couplings are given by
\beq
Y^{\nu}=V_{eL} \tilde{Y}^{\nu^\ast} V_{\nu_R}^{T}
\eeq
and the Lagrangian in this basis is
\bea
{\cal L}_{yuk}&=&(\bar{e}_L \bar{\mu}_L \bar{\tau}_L)
{\rm diag(m_e,m_{\mu},m_{\tau})} ({e}_R {\mu}_R {\tau}_R)^T
\nonumber \\
&+&(\bar{\nu_e}_L \bar{\nu_\mu}_L \bar{\nu_\tau}_L)Y^{\nu}v_2
(N_{R1} N_{R2} N_{R3} )^T
\nonumber \\
&+& (N_{R1} N_{R2} N_{R3} ){\rm diag(M_{1},M_{2},M_{3})} 
(N_{R1} N_{R2} N_{R3} )^T
+H.c.
\eea

\subsection{The See-Saw Mechanism and the MNS Matrix in this Basis}

The light effective left-handed Majorana
neutrino mass matrix in the above basis is
\beq
m_{LL}=v_2^2Y^{\nu}
{\rm diag(M_{1}^{-1},M_{2}^{-1},M_{3}^{-1})}  {Y^{\nu}}^T
\label{mLL}
\eeq
Having constructed the complex symmetric
light Majorana mass matrix it must then
be diagonalised by,
\beq
V_{\nu L}m_{LL}V_{\nu L}^{T}={\rm diag(|m_{1}|,|m_{2}|,|m_{3}|)}
\label{diag}
\eeq
where $V_{\nu L}$ is a unitary transformation and
the neutrino mass eigenvalues are real and positive.
The leptonic analogue of the CKM matrix
is the MNS matrix defined as \cite{MNS}
\beq
U_{MNS}=V_{eL}V_{\nu L}^{\dag},
\label{enu}
\eeq
where in the diagonal charged lepton basis $V_{eL}$ 
will only consist of a diagonal matrix of phases,
$V_{eL}=P_e$ corresponding to the charged lepton phase freedom,
\begin{equation}
\left(\begin{array}{c} e \\ \mu \\ \tau \end{array} \\ \right)_{L,R}
\rightarrow P_e
\left(\begin{array}{c} e \\ \mu \\ \tau \end{array} \\ \right)_{L,R}
\label{phases}
\end{equation}
where 
\beq
P_e=\left(\begin{array}{ccc}
e^{i\phi_{1}} & 0 & 0 \\
0 & e^{i\phi_{2}} & 0 \\
0 & 0 & e^{i\phi_{3}} \\
\end{array}\right)
\label{phases2}
\eeq
These transformations leave the
charged lepton masses real and positive,
and enable three phases to be removed 
from the unitary matrix $V_{\nu L}$, so that 
$U_{MNS}$ can be parameterized in terms of three mixing angles
$\theta_{ij}$ and three complex phases $\delta_{ij}$,
by regarding it as a product of three complex Euler rotations,
\begin{equation}
U_{MNS}=U_{23}U_{13}U_{12}
\label{MNS0}
\end{equation}
where
\begin{equation}
U_{23}=
\left(\begin{array}{ccc}
1 & 0 & 0 \\
0 & c_{23} & s_{23}e^{-i\delta_{23}} \\
0 & -s_{23}e^{i\delta_{23}} & c_{23} \\
\end{array}\right)
\label{MNS23}
\end{equation}
\begin{equation}
U_{13}=
\left(\begin{array}{ccc}
c_{13} & 0 & s_{13}e^{-i\delta_{13}} \\
0 & 1 & 0 \\
-s_{13}e^{i\delta_{13}} & 0 & c_{13} \\
\end{array}\right)
\label{MNS13}
\end{equation}
\begin{equation}
U_{12}=
\left(\begin{array}{ccc}
c_{12} & s_{12}e^{-i\delta_{12}} & 0 \\
-s_{12}e^{i\delta_{12}} & c_{12} & 0\\
0 & 0 & 1 \\
\end{array}\right)
\label{MNS12}
\end{equation}
where $c_{ij} = \cos\theta_{ij}$ and $s_{ij} = \sin\theta_{ij}$. 
The resulting MNS matrix is:

{\footnotesize 
\begin{equation}
\left(\begin{array}{ccc}
c_{12}c_{13} & s_{12}c_{13}e^{-i\delta_{12}} & s_{13}e^{-i\delta_{13}} \\
-s_{12}c_{23}e^{i\delta_{12}}-c_{12}s_{23}s_{13}
e^{i(\delta_{13}-\delta_{23})}
& c_{12}c_{23}-s_{12}s_{23}s_{13}e^{i(-\delta_{23}+\delta_{13}-\delta_{12})}
& s_{23}c_{13}e^{-i\delta_{23}} \\
s_{12}s_{23}e^{i(\delta_{23}+\delta_{12})}-c_{12}c_{23}s_{13}e^{i\delta_{13}}
& -c_{12}s_{23}e^{i\delta_{23}}
-s_{12}c_{23}s_{13}e^{i(\delta_{13}-\delta_{12})}
& c_{23}c_{13} \\
\end{array}\right)
\label{MNS}
\end{equation}
}

The Dirac phase which enters the CP odd part of
neutrino oscillation probabilities is given by
\beq
\delta = \delta_{13}-\delta_{23}-\delta_{12}.
\label{Dirac}
\eeq

\subsection{Leptogenesis in this Basis}

CP violation in the decay of the lightest right-handed neutrino
${N}_{R1}$ comes from the interference
between the tree-level and one-loop amplitudes
\cite{luty,lepto,covi,epsilon}. The CP asymmetries given by the
interference with the one-loop vertex amplitude are in the SM
\cite{luty,lepto}:
\begin{eqnarray}
\epsilon_1 &=& \frac{\Gamma ( {N}_{R1} \rightarrow {L_j} +
    H_2 ) - \Gamma({N}^\dagger_{R1} \rightarrow
    {L_j}^\dagger + H^\dagger_2)}
{\Gamma( {N}_{R1} \rightarrow {L_j} +
    H_2) + \Gamma({N}^\dagger_{R1} \rightarrow
    {L_j}^\dagger + H^\dagger_2) }
\nonumber \\
    &=&\frac{1}{8\pi(Y_{\nu}^\dagger Y_{\nu})_{11}}
\sum_{i\neq 1} 
Im \left(  \left[ (Y_{\nu}^\dagger Y_{\nu})_{1i}\right]^2 \right) 
\left( f(\frac{M_{1}^2}{M_{i}^2})+g(\frac{M_{1}^2}{M_{i}^2}) \right)
\end{eqnarray}   
where
\begin{equation}
f(x) = \sqrt{x} \left[ 1 - (1+x) \ln \left( \frac{1+x}{x} \right)
\right], \ \ \ g(x) = \frac{\sqrt{x}}{1-x} ,
\end{equation}
where $f(x)$ arises from the interference between the tree level decay
and the vertex correction, while $g(x)$ is due to
the interference with the absorptive part of the one-loop self-energy,
which can in principle be much larger if the right-handed neutrinos are 
almost degenerate \cite{covi,epsilon}. 
Assuming that $M_1\ll M_2 \ll M_3$, 
we have approximately \cite{lepto2},
\begin{equation}
\epsilon_1 \approx -\frac{3}{16\pi(Y_{\nu}^\dagger Y_{\nu})_{11}}
\sum_{i\neq 1} 
Im \left(  \left[ (Y_{\nu}^\dagger Y_{\nu})_{1i}\right]^2 \right) 
\left(\frac{M_{1}}{M_{i}} \right)
\label{BP}
\end{equation}   
In the Supersymmetric SM the result for $\epsilon_1$ is twice as large
as in Eq.\ref{BP} due to the extra SUSY degrees of freedom
in the diagrams.

The lepton asymmetry $Y_L$ of the universe created by this mechanism 
can be written as 
\begin{equation}
Y_L = d \frac{\epsilon_1}{g^*}
\label{defyl}
\end{equation}
where $\epsilon_1$ has been defined above, $g^*$ counts the effective 
number of degrees of freedom, for the SM $g^* = 106.75$ while for the 
Supersymmetric SM $g^* = 228.75$ \cite{kolb} and $d$ is the dilution 
factor which takes into account the washout effect produced by inverse 
decay and lepton number violating scattering. To calculate $d$ one has 
to solve, in principle, the full Boltzman equations, which can be done 
numerically \cite{luty,lepto2}. 

However, many authors, for examples see \cite{kolb}, have used  
simple approximated solutions to the Boltzman equations expressed 
as \cite{kolb}
\begin{equation}
d=\frac{0.24}{k(lnk)^{3/5}} \hskip10mm 10 \le k \le 10^6 
\label{dold1}
\end{equation}   
\begin{equation}
d=\frac{1}{2k} \hskip10mm 1 \le k \le 10 
\label{dold2}
\end{equation}   
\begin{equation}
d=1 \hskip10mm 0 \le k \le 1 
\label{dold3}
\end{equation}   

Recently, Nielson and Takanishi \cite{yasutaka} suggested a slight 
modification of eqs (\ref{dold1})-(\ref{dold3}), namely, 

\begin{equation}
d=\frac{0.3}{k(lnk)^{3/5}} \hskip10mm 10 \le k \le 10^6 
\label{dtak1}
\end{equation}   
\begin{equation}
d=\frac{1}{2\sqrt{k^2+9}} \hskip10mm  k \le 10 \nonumber 
\label{dtak2}
\end{equation}   
Here the parameter $k$ is given by,
\begin{equation}
k = \frac{M_P}{1.7 \times 8 \pi \sqrt{g^*}}
\frac{(Y_{\nu}^{\dagger}Y_{\nu})_{11}}{M_1}
\end{equation}
where $M_P$ is the Planck mass. 
Physically $k\sim 1$ represents the desirable region where
the couplings of the right-handed neutrinos are sufficiently strong for 
them to be copiously 
produced from particles in the thermal bath, but sufficiently 
weak for them to decay satisfying the out-of-equilibrium condition.

We have compared eqs (\ref{dtak1})-(\ref{dtak2}) to the full numerical 
solution to the Boltzman equations as plotted in fig. 6 
of \cite{FitExact}. Buchm\"uller and Pl\"umacher \cite{FitExact} 
use as a variable 

\begin{equation}
{\tilde m}_1 = v^2 
\frac{(Y_{\nu}^{\dagger}Y_{\nu})_{11}}{M_1}
\simeq 1.1 \times 10^{-3} k {\ \rm eV}. 
\label{defmtilde}
\end{equation}

\begin{figure}
\setlength{\unitlength}{1mm}
\begin{picture}(0,80)
\put(10,0)
{\mbox{\epsfig{file=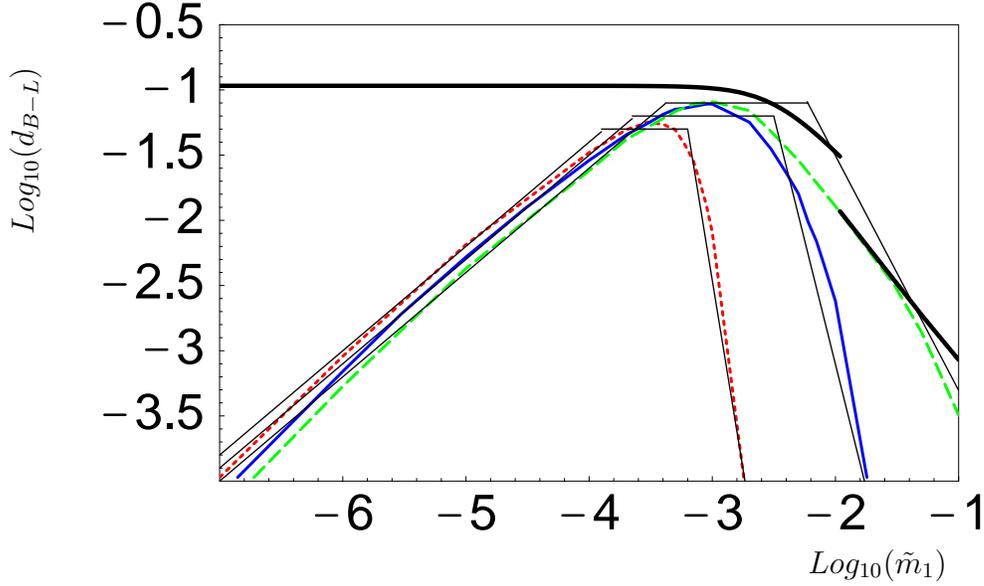,height=12.0cm,width=12.0cm}}}
\end{picture}
{\vskip-70mm\hskip2mm \begin{rotate}{90}
$Log_{10}(d_{B-L})$
\end{rotate}
\vskip40mm\hskip105mm $Log_{10}({\tilde m}_1)$ 
\vskip2mm}
\caption[allfig]{The logarithm of the dilution function $d_{B-L}$ 
versus the logarithm of ${\tilde m}_1$ for different values of the lightest 
right-handed neutrino mass $M_1 = 10^8$, $10^{10}$, $10^{12}$ GeV
(from right to left). The thick line is the solution to 
Eqs.(\ref{dtak1})-(\ref{dtak2}). The dashed, medium full,
dotted curves represent $d_{B-L}$ for
$M_1 = 10^8$, $10^{10}$, $10^{12}$ GeV extracted from Fig.(6) of 
\cite{FitExact}, based on an exact numerical solution of the 
Boltzman equation.
The thin solid curves with plateau regions
are the approximately fitted values of $d_{B-L}$ 
following Eqs. (\ref{dnew1})-(\ref{dnew3}).}
        \label{fig:pluemi}
\end{figure}

The result of this comparison is shown in Fig. 
\ref{fig:pluemi}. As one can see, the approximate formulas, 
eqs (\ref{dtak1})-(\ref{dtak2}), are a valid approximation for values 
of $k\sim 1$ and for small values of $M_1 \le 10^{10}$ GeV. 
For smaller values of $k\ll 1$ or 
${\tilde m}_1\ll 10^{-3} {\ \rm eV}$ 
the approximation formulas are clearly not valid 
since they do not take into account the production suppression
apparent in the full treatment using the Boltzmann equations.
For larger values of $k\gg 1$ or 
${\tilde m}_1\gg 10^{-3} {\ \rm eV}$ 
(and larger values of $M_1$)
the approximation formulas are also clearly not valid 
since they do not take into account the steep suppression
due to the out-of-equilibrium condition being violated
which is again apparent in the full treatment using the Boltzmann equations. 
From Fig. \ref{fig:pluemi} it is obvious that in this case the 
analytic approximation seriously underestimates the suppression 
of $d$ by orders of magnitude.

For this reason we have devised a purely empirical fit formula for 
the exact solution to the Boltzman equations, which can be written 
as,

\bea
(a) \hskip2mm Log_{10}(d_{B-L}) &=& 0.8 * Log_{10}({\tilde m}_1) + 1.7 
              + 0.05 * Log_{10}(M_1^{10}) \label{dnew1} \\
(b) \hskip2mm Log_{10}(d_{B-L}) &=& -1.2 
             - 0.05 * Log_{10}(M_1^{10}) \label{dnew2}\\
(c) \hskip2mm Log_{10}(d_{B-L}) &=&-(3.8+Log_{10}(M_1^{10}))
               *(Log_{10}({\tilde m}_1)+2) \label{dnew3}\\ \nonumber
            &-&(5.4 - \frac{2}{3} *Log_{10}(M_1))^2 -\frac{3}{2}
\eea
where $M_1^{10} = M_1/{\rm 10^{10} GeV}$, ${\tilde m}_1$ is in 
units of [eV]. In implementing this fit it is 
always the smallest of (a)-(c) which is taken.
\footnote{We fit $d_{B-L}$ since 
authors of \cite{FitExact} plot $Y_{B-L}$, where 
$d_{B-L}$ is related to $d$ via $d=(1-\alpha) d_{B-L}$, where $\alpha$ 
is defined in Eq.\ref{alpha}.}
The results of this fit are superimposed onto the 
exact curves taken from \cite{FitExact} in Fig. \ref{fig:pluemi}. 

\begin{figure}
\setlength{\unitlength}{1mm}
\begin{picture}(0,55)
\put(5,-5)
{\mbox{\epsfig{file=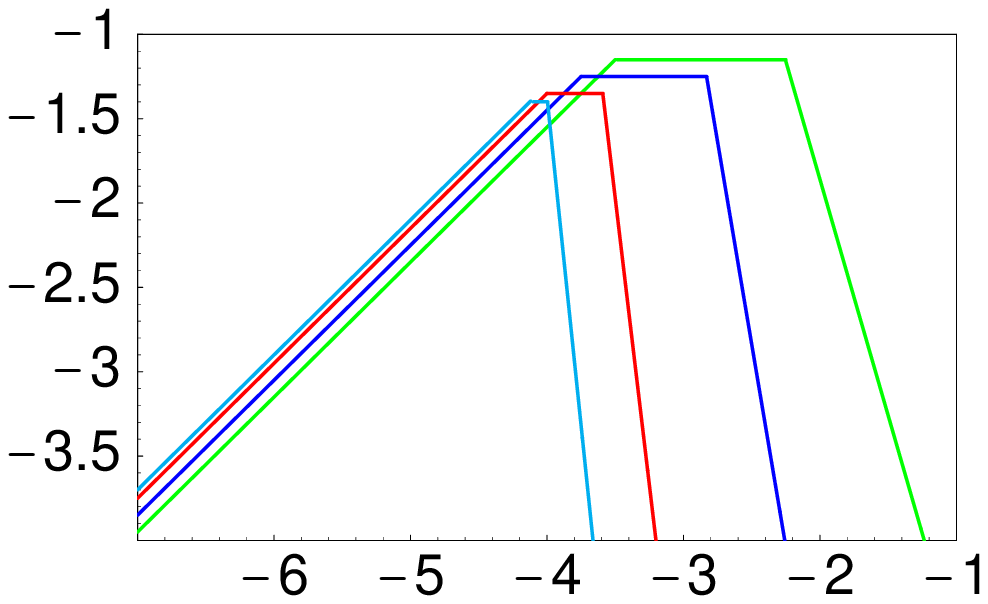,height=8.0cm,width=8.0cm}}}
\end{picture}
{\vskip-40mm\hskip2mm \begin{rotate}{90}
$Log_{10}(d_{B-L})$
\end{rotate}
\vskip23mm\hskip60mm $Log_{10}({\tilde m}_1)$ 
\vskip2mm}
\caption[allfig]{The approximate solutions of the fit function for $d_{B-L}$ 
from Eqs.(\ref{dnew1})-(\ref{dnew3})) for (from right to left) 
$M_1 = 10^9,10^{11},10^{13},10^{14}$ GeV. 
For further discussion see caption in Fig.\ref{fig:pluemi}.}
        \label{fig:pluemipred}
\end{figure}

Obviously eqs. (\ref{dnew1})- (\ref{dnew3}) reproduce the exact 
results considerably better than eqs (\ref{dtak1})-(\ref{dtak2}).
However, in \cite{FitExact} the 
authors assumed that right-handed neutrinos are hierarchical. Also 
we have to assume that for small values of $\epsilon_1$ the dilution 
function does not depend on $\epsilon_1$. Thus, eqs (\ref{dnew1})-
(\ref{dnew3}) are still approximate, and may 
not be valid if these conditions are violated. Moreover, 
for large values of $M_1$, say $M_1 \sim 10^{14}$, we have to rely on an 
extrapolation beyond values of $M_1$ used in the fit.

Note also that since in the SUSY SM both $\epsilon_1$ 
and $g^*$ are twice as large as in the SM, the two effects
tend to cancel in the estimate of $Y_L$. Also the approximations
for $d$ over the above range of ${\tilde m}_1$ are valid for either the
SM or the SUSY SM \cite{kolb}. 
Therefore the results we present are approximately
valid for either the SM or the SUSY SM, although for definiteness
we consider the SM from now on.

Due to sphaleron effects $Y_L$ finally is related to $Y_B$ approximately 
via \cite{har90}
\begin{equation}
Y_B = \frac{\alpha}{\alpha-1} Y_L
\label{defyb}
\end{equation}
where
\begin{equation}
\alpha = \frac{8N_F + 4N_H}{22N_F+13N_H}
\label{alpha}
\end{equation}
Here $N_F$ is the number of families and $N_H$ the number of Higgs 
doublets. In the SM $\alpha \simeq 1/3$. Experimentally $Y_B$ is expected to 
be in the range $Y_B = (n_B-n_{\bar B})/s \sim (0.5-1) \times 10^{-10}$ 
\cite{lepto,epsilon}.

\section{Analytic Estimates}

\subsection{MNS Parameters from SRHND}

In the basis used in this paper where the charged leptons
are diagonal, and the right-handed neutrinos are diagonal,
we write the neutrino Yukawa matrix as
\begin{equation}
Y_{\nu}=
\left( \begin{array}{ccc}
a' & a & {d}\\
b' & b & {e}\\
c' & c & {f}
\end{array}
\right)
\label{Y1}
\end{equation}
where the LR notation means that the second and third columns
of $Y_{\nu}$ correspond to the second and third right-handed neutrinos.
We use the phase freedom of the charged lepton masses 
in Eq.\ref{phases} to 
make the couplings to the third right-handed neutrino
$d,e,f$ real and positive, leaving $a,b,c,a',b',c'$ complex.

We write the diagonal (real, positive) Majorana masses in this basis as
\begin{equation}
M_{RR}=
\left( \begin{array}{ccc}
X' & 0 & 0    \\
0 & X & 0 \\
0 & 0 & Y
\end{array}
\right) 
\end{equation}
Then using the see-saw formula for the light effective Majorana
mass matrix $m_{LL}=v_2^2Y^{\nu}M_{RR}^{-1}{Y^{\nu}}^T$
(valid for complex couplings) we find
the symmetric matrix,
\begin{equation}
m_{LL}
=
\left( \begin{array}{ccc}
\frac{a'^2}{X'}+\frac{a^2}{X}+\frac{d^2}{Y}
& \frac{a'b'}{X'}+\frac{ab}{X}+ \frac{de}{Y}
& \frac{a'c'}{X'}+\frac{ac}{X}+\frac{df}{Y}    \\
.
& \frac{b'^2}{X'}+\frac{b^2}{X}+\frac{e^2}{Y} 
& \frac{b'c'}{X'}+\frac{bc}{X}+\frac{ef}{Y}    \\
.
& .
& \frac{c'^2}{X'}+\frac{c^2}{X}+\frac{f^2}{Y} 
\end{array}
\right)
\end{equation}

So far the discussion is completely general.
In order to account for the atmospheric
and solar neutrino data many models have been proposed \cite{SelModels} 
based on the see-saw mechanism \cite{seesaw}.
One question which is common to all these models is
how to arrange for a large mixing angle involving the
second and third generation of neutrinos, without destroying the
hierarchy of mass splittings necessary to account for
the solar and atmospheric data.
Assuming $\theta_{23}\sim \pi/4$ one might expect
two similar eigenvalues $m_2 \sim m_3$, and then a hierarchy
of neutrino masses seems rather unnatural.

For our analytic estimates, we assume for simplicity that the 
first right-handed neutrino $X'$
contributions are insignificant compared to the second right-handed
neutrino $X$ contributions,
\begin{equation}
\frac{|a'+b'+c'|^2}{X'}\ll
\frac{|a+b+c|^2}{X}
\end{equation}
Then one way to achieve a natural hierarchy is to 
suppose that the third right-handed neutrino contributions 
are much greater than the second right-handed neutrino contributions
in the 23 block of $m_{LL}$ \cite{SRHND2},
\begin{equation}
\frac{(e^2,ef,f^2)}{Y}\gg
\frac{|a+b+c|^2}{X}
\label{SRHND}
\end{equation}

This implies an approximately vanishing 23 subdeterminant,
\begin{equation}
det[m_{LL}]_{23}=
\left( \frac{e^2}{Y} +\frac{b^2}{X}\right)
\left( \frac{f^2}{Y} +\frac{c^2}{X}\right)
-\left( \frac{ef}{Y} +\frac{bc}{X}\right) ^2\approx 0
\label{det1}
\end{equation}
and hence
\begin{equation}
{m_{2}}/{m_{3}}\ll 1
\end{equation}
Thus the assumption in Eq.\ref{SRHND} that the right-handed neutrino $Y$
gives the dominant contribution to the 23 block of
$m_{LL}$ naturally leads to a neutrino mass hierarchy.
This mechanism is called single right-handed neutrino dominance
(SRHND)\cite{SRHND}. In the limit that only a single right
handed neutrino contributes the determinant clearly exactly vanishes and we 
have $m_{2}=0$ exactly. However the sub-dominant contributions
from the right-handed neutrino $X$ will give a small finite
mass $m_{2}\neq 0$ as required by the MSW solution to the
solar neutrino problem. 

Assuming SRHND as discussed above,
we may obtain a simple estimate for the third
neutrino mass:
\begin{equation}
m_{3}\approx v_2^2\frac{(d^2+e^2+f^2)}{Y}
\label{m3}
\end{equation}
Note that $m_{1,2}$ are determined by 
parameters associated with the subdominant
right-handed neutrinos and so are naturally smaller.
Given the SRHND assumption in Eq.\ref{SRHND}
we see that we have generated a hierarchical spectrum
$|m_{1,2}|\ll |m_3|$.

In order to obtain the MNS parameters
we must diagonalise $m_{LL}$ as in Eq.\ref{diag},
\beq
V_{\nu L}m_{LL}V_{\nu L}^{T}={\rm diag(|m_{1}|,|m_{2}|,|m_{3}|)}
\label{diag2}
\eeq
where we write $V_{\nu L}$ as a product of complex
Euler rotations of the form of
Eqs.\ref{MNS0},\ref{MNS23},\ref{MNS13},\ref{MNS12}, 
together with diagonal phase matrix  
\beq
P_{\nu}=\left(\begin{array}{ccc}
e^{i\tilde{\phi}_{1}/2} & 0 & 0 \\
0 & e^{i\tilde{\phi}_{2}/2} & 0 \\
0 & 0 & e^{i\tilde{\phi}_{3}/2} \\
\end{array}\right)
\label{phases3}
\eeq
which is required to remove the phases in 
$m_i=|m_i|e^{i\tilde{\phi}_i}$,
\beq
V_{\nu L}=P_{\nu}^\dagger \tilde{U}_{12}^\dagger 
\tilde{U}_{13}^\dagger \tilde{U}_{23}^\dagger 
\eeq	
Thus $V_{\nu L}$ contains the 3 angles and 6 phases of a general
unitary matrix.
However in the basis where we have
chosen the couplings $d,e,f$ to be real, $m_3$ is given in Eq.\ref{m3}
and $\tilde{\phi}_3$ is zero to leading order.

In order to bring the MNS matrix into the form in Eq.\ref{MNS}, additional
charged lepton phase rotations are required as in
Eqs.\ref{enu}, so that we have finally
\beq
U_{MNS}=P_{e} \tilde{U}_{23} 
\tilde{U}_{13} \tilde{U}_{12} P_{\nu}
\eeq	
where $P_{e}$ is a diagonal matrix of phases as in
Eq.\ref{phases2}. Note that the angles involved in the
$\tilde{U}_{ij}$ are the same as those in the ${U}_{ij}$
in Eq.\ref{MNS0}, $\tilde{\theta}_{ij}={\theta}_{ij}$,
but the phases will be different,
$\tilde{\delta}_{ij}\neq {\delta}_{ij}$, due to the
non-zero phases in $P_e , P_{\nu}$.

Since the couplings $d,e,f$ are real,
we find that the previous estimates based on SRHND are
still valid \cite{SRHND2}  
\begin{equation}
\tan \theta_{23} \approx {\frac{e}{f}}, 
\end{equation}
\begin{equation}
\tan \theta_{13} \approx {\frac{d}{\sqrt{e^2+f^2}}}
\end{equation}
where the associated phases are approximately zero
\beq
\tilde{\delta}_{23}\approx \tilde{\delta}_{13} \approx 0
\eeq
By a suitable choice of parameters $e=f\gg d$ 
it is possible to
have maximal $\theta_{23}$ suitable for atmospheric oscillations, 
while maintaining
a small $\theta_{13}$ consistent with the CHOOZ constraint \cite{chooz}.

To determine $U_{12}$ is quite complicated in general,
but in the physically interesting cases where $\theta_{12}$
is near maximal $\theta_{12}\approx \pi/4$ 
we find the simple analytical results
\begin{equation}
\tan \theta_{12} \approx {\sqrt{2}\frac{|a|}{|b-c|}}
\end{equation}
\begin{equation}
\tilde{\delta}_{12} \approx \phi_{b-c} - \phi_a
\label{DiracSim}
\end{equation}
where $\phi_{b-c} =arg(b-c)$ and $\phi_a =arg(a)$.
In the simple example that the phases in $P_e , P_{\nu}$ are zero,
the observable Dirac phase in Eq.\ref{Dirac} is given in
Eq.\ref{DiracSim}.
In general the Dirac phase will involve a more complicated
combination of phases.

\subsection{Leptogenesis in SRHND}

In leptogenesis it is generally the lightest right-handed
neutrino which decays to produce lepton number, where
we use the notation that $M_1$ is the lightest right-handed neutrino, 
$M_3$ is the heaviest right-handed neutrino and we
assume $M_1\ll M_2 \ll M_3$.
In the notation of the previous subsection where $Y$
is the dominant right-handed neutrino
there are two physically distinct cases to consider:

(a) $Y\ll X \ll X'$  (i.e. $Y=M_1$, $X=M_2$, $X'=M_3$)

(b) $X'\ll X \ll Y$  (i.e. $X'=M_1$, $X=M_2$, $Y=M_3$)

In other words the dominant right-handed neutrino may either
be (a) the lightest, or (b) the heaviest right-handed neutrino,
and both cases must be considered.

It is also worth emphasising that there is no generation
ordering implied by the results in the previous
subsection (or those in \cite{SRHND},\cite{SRHND2}).
In other words the dominant right-handed neutrino $Y$ 
may be associated with the third, second or first generation,
by a simple reordering of the columns of $Y_{\nu}$. 
Due to the hierarchy of charged lepton masses,
it is meaningful to associate 
the first row of $Y_{\nu}$ with the first generation, 
the second row of $Y_{\nu}$ with the second generation, 
and the third row of $Y_{\nu}$ with the third generation.
However the physical
neutrino mass matrix $m_{LL}$ is invariant under the
operation of exchanging the {\em columns} of $Y_{\nu}$,
along with the ordering of the right-handed neutrinos
in $M_{RR}$, so the SRHND results apply quite generally
to all generation orderings of the right-handed neutrinos
\cite{SRHND}, \cite{SRHND2}.
Physically if the Yukawa couplings $e,f$ are of order unity,
then it may be natural to associate $Y$ with the third generation.
However if the couplings $e,f\ll 1$ then it may be more natural
to associate $Y$ with the second generation, and re-order
the matrices by
interchanging of the second and third right-handed neutrinos in $Y_{\nu}$
and $M_{RR}$. 

Returning to the leptogenesis asymmetry parameter in Eq.\ref{BP},
for case (a), where the dominant right-handed neutrino mass $Y$ 
is the lightest,
using the SRHND results of the previous
subsection, we find
\begin{equation}
\epsilon_1^{(a)} \approx -\frac{3}{32\pi}
\left(\frac{Y}{X} \right)
\sin(2\phi_{b+c})|b+c|^2
\label{a}
\end{equation}   
while for case (b), where the dominant right-handed neutrino mass $Y$ 
is the heaviest, we find
\begin{equation}
\epsilon_1^{(b)} \approx \frac{3}{16\pi}
\left(\frac{X'}{Y} \right)
\sin(2\phi_{b'+c'})e^2\frac{|b'+c'|^2}{|a'|^2+|b'|^2+|c'|^2}
\label{b}
\end{equation}   
where $\phi_{b+c} =arg(b+c)$ and $\phi_{b'+c'} =arg(b'+c')$,
and we have used the fact that $m_2\ll m_3$ in obtaining Eq.\ref{b}.
\footnote 
{It is also apparent that the phases which are relevant for leptogenesis
in both cases are not identical to the Dirac phase which even in the
simple example that the phases in $P_e , P_{\nu}$ are zero, is given as
in Eq.\ref{Dirac} as $\delta \approx \phi_a - \phi_{b-c}$. 
In general the Dirac phase will involve a more complicated
combination of phases still. }

Are these values of $\epsilon_1$ of the correct order of magnitude?
We may use $m_3\sim 5\times 10^{-2}$eV 
and $m_{3}\approx v_2^2\frac{(2e^2)}{Y}$
in Eq.\ref{m3},
and the crude order of magnitude approximation
for $m_2 \sim |b-c|^2v_2^2/M_2$, to obtain
\bea
\epsilon_1^{(a)} 
& \sim & 
\sin(2\phi_{b+c}) 10^{-5}\left(\frac{m_2}{m_3} \right)
\left(\frac{Y}{10^{11}GeV} \right)
\label{a1}
\eea   
\bea
\epsilon_1^{(b)} 
& \sim &  
\sin(2\phi_{b'+c'})
10^{-6}
\left(\frac{X'}{10^{10}GeV} \right)
\label{b1}
\eea
The results in Eqs.\ref{a1}, \ref{b1} express $\epsilon_1$
in terms of the lightest right-handed neutrino mass in each case.
Since $\epsilon_1^{(a)}$ is suppressed relative to $\epsilon_1^{(b)}$
by a factor of $m_2/m_3$ (which should be $m_2/m_3 <0.1$), this implies that 
the lightest right-handed neutrino mass must be at least an order of magnitude
larger in case (a) than in case (b). 
\footnote{Note that since the dominant right-handed
neutrino mass is given by $Y\sim e^2 5.10^{14}$ GeV,
case (a) requires 
$e\ll 1$, whereas for case (b) it is consistent with $e \sim 1$
providing there is a sufficiently large hierarchy in the
right-handed neutrino sector.
This means that in case (a) the dominant right-handed
neutrino cannot be associated with the third family, whereas
in case (b) it may be.} 

To understand which of the two cases (a) or (b) is more promising
from the point of view of leptogenesis it is important to estimate
the parameter ${\tilde m}_1$ in Eq.\ref{defmtilde}
which controls the dilution factor as shown in Figs.
\ref{fig:pluemi},\ref{fig:pluemipred}. 
From Eqs.\ref{defmtilde},\ref{Y1},
\begin{equation}
{\tilde m}_1^{(a)} \simeq v_2^2 \frac{(|d|^2 + |e|^2 + |f|^2)}{Y}
\label{m1a}
\end{equation}
\begin{equation}
{\tilde m}_1^{(b)}\simeq v_2^2 \frac{(|a'|^2 + |b'|^2 + |c'|^2)}{X'}
\label{m1b}
\end{equation}

In case (a), where the dominant right-handed neutrino is the lightest 
one, the parameter ${\tilde m}_1^{(a)}$ in Eq.\ref{m1a} is approximately 
equal to the physical mass of the heaviest neutrino in Eq.\ref{m3} 
which is measured by Super-Kamiokande. Thus for these models 
${\tilde m}_1^{(a)} \sim m_{3} \sim 5\times 10^{-2}$ eV
which is generally beyond the plateau regions in 
Figs.\ref{fig:pluemi}, \ref{fig:pluemipred}, and this leads to the
requirement that $Y \sim M_1 < 10^9$ GeV
and a strong dilution suppression $d\ll 1$.
However, according to Eq.\ref{a1}, $Y \sim M_1 < 10^9$ GeV
leads to values of 
$\epsilon_1^{(a)} < 10^{-8}$ which, 
when combined with the 
dilution suppression $d\ll 1$, implies from Eq.\ref{defyl} $Y_B\ll 10^{-10}$ 
well below the observed value.

In case (b), on the other hand, where the dominant right-handed neutrino 
is the heaviest one, there is no association of ${\tilde m}_1^{(b)}$ in 
Eq.\ref{m1b} with a physical neutrino mass and so this parameter may 
in principle take 
smaller values closer to the plateau regions,
leading to only a mild
dilution suppression $d\simlt 0.1$
for a range of lightest
right-handed neutrino mass $X'$.
Furthermore, as we already remarked, by comparing Eqs.\ref{a1}, \ref{b1}
we see that value of $\epsilon_1^{(b)}$ is larger by an order
of magnitude than $\epsilon_1^{(a)}$.
For example if we choose $X' \sim M_1 \simlt 10^9$ GeV,
consistent with the gravitino constraint on the 
reheating temperature $T_R\simlt 10^9 {\rm GeV}$ \cite{gravitino},
we find $\epsilon_1^{(b)} \simlt 10^{-7}$ which, assuming a mild
dilution suppression $d\simlt 0.1$, implies from Eq.\ref{defyl}
$Y_B \simlt 10^{-10}$ which is just about acceptable.

We shall later present specific examples with 
detailed numerical results
which support the conclusion that leptogenesis prefers
case (b) where the dominant right-handed neutrino is the
heaviest one, at least according to the standard hot big
bang picture, ignoring effects of inflation.

\section{Numerical Approach to $U(1)$ Family Symmetry Models}

Our numerical results are based on the SRHND models 
\cite{SRHND},\cite{SRHND2}, with 
a $U(1)$ family symmetry. 
The idea of such a symmetry is that the three families of leptons
are assigned different $U(1)$ charges, and these different charges
then control the degree of suppression of the operators responsible
for the Yukawa couplings, leading to Yukawa matrices with 
a hiearchy of entries, and approximate ``texture'' zeroes \cite{flavour}.
As usual it is assumed that the $U(1)$ 
is slightly broken by the VEVs of some 
fields $\theta , \bar{\theta}$ 
which are singlets under the standard model gauge group, but
which have vector-like charges $\pm 1$ under the $U(1)$ flavour
symmetry. The $U(1)$ breaking 
scale is set by $<\theta >=<\bar{\theta} >$. Additional exotic vector 
matter with mass $M_V$ allows an expansion parameter $\lambda$ to be 
generated by a Froggatt-Nielsen mechanism \cite{flavour},
\begin{equation}
\frac{<\theta >}{M_V}=\frac{<\bar{\theta} >}{M_V}= \lambda \approx 0.22
\label{expansion}
\end{equation}
where the numerical value of $\lambda$ is motivated by the size of 
the Cabibbo angle. Small Yukawa couplings are generated effectively 
from higher dimension non-renormalisable operators corresponding 
to insertions of $\theta$ and $\bar{\theta}$ fields and hence 
to powers of the expansion parameter in Eq.\ref{expansion}. The number 
of powers of the expansion parameter is controlled by the $U(1)$ charge 
of the particular operator. The lepton doublets, neutrino singlets, 
Higgs doublet and Higgs singlet relevant to the construction of 
neutrino mass matrices are assigned $U(1)$ charges $l_i$, $n_p$, 
$h_u=0$ and $\sigma$. From this starting  point one may then
generate the neutrino Yukawa matrices as in \cite{SRHND}. 
The neutrino Dirac Yukawa matrix is 
\begin{equation}
\tilde{Y}^{\nu} = 
\left( \begin{array}{ccc}
a_{11} \lambda^{|l_1+n_1|} & a_{12} \lambda^{|l_1+n_2|} 
&  a_{13} \lambda^{|l_1+n_3|} \\ 
a_{21} \lambda^{|l_2+n_1|} & a_{22} \lambda^{|l_2+n_2|} 
&  a_{23} \lambda^{|l_2+n_3|} \\ 
a_{31} \lambda^{|l_3+n_1|} & a_{32} \lambda^{|l_3+n_2|} 
&  a_{33} \lambda^{|l_3+n_3|} 
\end{array}
\right)
\label{Y2}
\end{equation}
where Eq.\ref{Y2} may be identified with Eq.\ref{Y1}.
The heavy Majorana matrix is
\begin{equation}
\tilde{Y}_{RR} = 
\left( \begin{array}{ccc}
A_{11} \bar{\lambda}^{|2n_1+\sigma|} 
& A_{12} \bar{\lambda}^{|n_1+n_2+\sigma|} 
&  A_{13} \bar{\lambda}^{|n_1+n_3+\sigma|} \\ 
A_{12} \bar{\lambda}^{|n_2+n_1+\sigma|} 
& A_{22} \bar{\lambda}^{|2n_2+\sigma|} 
&  A_{23} \bar{\lambda}^{|n_2+n_3+\sigma|} \\ 
A_{13} \bar{\lambda}^{|n_3+n_1+\sigma|} 
& A_{23} \bar{\lambda}^{|n_3+n_2+\sigma|} 
&  A_{33} \bar{\lambda}^{|2n_3+\sigma|} 
\end{array}
\right)
\label{MRR}
\end{equation}
where $A_{ij}$ and $a_{ij}$ are undetermined coefficients,
and $\bar{\lambda}$ is an independent expansion parameter relevant
for the right-handed neutrino sector.
\footnote{We are grateful to G.Ross for emphasising that the
right-handed neutrino sector is controlled by an independent
expansion parameter.}

The neutrino Yukawa matrices are generated in a
particular basis defined by the $U(1)$ family symmetry.
This corresponds to the starting basis 
defined by tildes in section 2, and numerically we follow the procedure 
to go to the diagonal right-handed neutrino basis, as outlined there.
Note that we assume as an approximation that the charged lepton matrix
is diagonal with positive eigenvalues in the starting basis.
In practice this may be approximately achieved by a suitable choice
of right-handed lepton $U(1)$ family charges, as discussed
elsewhere \cite{SRHND},\cite{SRHND2}. 

In our numerical analysis we take account of the fact that
the theory does not determine the complex coefficients $A_{ij}$ and 
$a_{ij}$ which one has to choose in some range. This is not a special 
feature of the SRHND models, which we are focussing on in this paper, 
but a limitation of texture models based on a $U(1)$ family 
symmetry. Usually one simply assumes that the unknown 
coefficients are of order ${\cal O}(1)$ and, therefore, the structure 
in the Yukawa matrices is given by the expansion parameter rather 
than the coefficients.
Our approach to this problem is to scan over the 
unknown coefficients randomly and to construct distributions for the 
various observables of interest. This way we are able to determine 
distributions for masses and mixings of a given model. 
Given the statistical nature of our approach, one question comes 
immediately to mind: What is the correct range of values one should 
choose for the coefficients? Lacking any theoretical background we 
have chosen for the coefficients the interval

\begin{equation}
a_{ij}, A_{ij} \Rightarrow [\sqrt{\lambda},1/\sqrt{\lambda}] \times 
e^{i \phi_{ij}}, \hskip10mm \phi_{ij} \Rightarrow [0,2\pi]
\end{equation}

It should be noted, that this choice is {\it the minimum requirement 
for texture models to be sensible}, simply because any larger variation 
in the coefficients would destroy the texture one originally assumed 
to be the dominant feature of the mass matrices of interest. 

A word of caution might be in order. Obviously the distributions which 
we calculate depend on our choice for $a_{ij}, A_{ij}$. Lacking further 
theoretical support for our choice, we can not evaluate the success of a 
given model in terms of confidence intervals. Instead our method is more 
minimalistic. We will consider a model to be a ``good'' model, if the 
main body of the distribution in a given observable coincides with or is 
close to the experimentally preferred value. Clearly, a model which fails 
even our simplistic test will fail even more badly under a more sophisticated 
numerical analysis. We would like to stress, however, that although 
the width of the peaks and the detailed shape of the distributions 
change under a change of the range of the coefficients, the {\it position 
of the peaks} remains nearly invariant. 

In order to be able to compute the expectations for the leptogenesis 
``observable'' $\epsilon_1$ in the different models, our current computation 
goes beyond the one we discussed in a previous paper \cite{discr} in 
allowing the coefficients $a_{ij}, A_{ij}$ to be complex. Since we do 
not have a theory of phases, we decided to choose the $\phi_{ij}$ in 
the full interval $[0,2\pi]$. In other words, since we do not know about 
any mechanism suppressing phases effectively in the Yukawa couplings, we 
simply expect that all phases should be large. 

So, our numerical procedure may be summarised as follows.
First select a particular flavour model defined by a choice 
of $U(1)$ charges. Second select randomly a set of complex
coefficients $a_{ij}$ and $A_{ij}$.
Third diagonalise the right-handed neutrino mass matrix
to yield positive eigenvalues, and express the Dirac Yukawa
matrix in this basis, as discussed in section 2.1.
Fourth calculate the see-saw matrix $m_{LL}$
and hence the physical neutrino masses and the MNS angles and
three phases as discussed in section 2.2.
Fifth calculate the
leptogenesis parameters $\epsilon_1$ and $Y_B$
as discussed in section 2.3.
Then the whole procedure is repeated for a different set of 
randomly chosen complex coefficients $a_{ij}$ and $A_{ij}$,
and the results are binned to build up distributions
of the observable quantities. In the figures we show in the 
following we use random sets of $10^8$ matrices for each of 
the distributions shown. 
Finally a different model is selected corresponding to a different
set of $U(1)$ charges and the whole procedure is repeated for the
new model. We disregard the effect of renormalisation group
radiative corrections in going from high energy to low energy,
which has been demonstrated to be of the order of a few per cent
for SRHND models \cite{SRHNDRGE}

\section{Leptogenesis Decoupling}

In this section we will discuss leptogenesis {\em decoupling}, 
namely, the fact that the leptogenesis observable $\epsilon_1$ 
can take any value independent of the low energy observables, 
i.e. masses and mixings. Unfortunately this means that measurements 
of the solar angle or the MNS phase for example does not tell us 
anything about leptogenesis. On the other hand the results in this 
section also demonstrate another aspect of leptogenesis, 
namely that it can be used to resolve the ambiguity between different 
models which all lead to very similar predictions for low energy 
neutrino observables. In this way leptogenesis provides 
information about the high energy theory which would be impossible
to determine by the measurement of low energy observables alone.

We will defer the discussion of $Y_B$ until the next section and 
concentrate here only on the calculation of $\epsilon_1$, since 
the conversion of the CP asymmetry parameter to $Y_B$ depends highly 
on the assumed thermal history of the universe, whereas the calculation 
of $\epsilon_1$ is theoretically clean.

In Table \ref{FC} we give four examples of models based on different 
choices of flavour charges. 
For simplicity, we start by assuming that the expansion
parameter in the right-handed neutrino sector
is equal to the Wolfenstein parameter $\bar{\lambda}= \lambda$,
as was assumed in \cite{SRHND2}.
Model FC1 was discussed analytically in \cite{SRHND2},
where it is seen that it yields a heavy Majorana matrix with
an off-diagonal structure in the $U(1)$ charge basis.
It satisfies the SRHND conditions, and has $a\sim b,c$
and so leads to the LMA solution.
FC2 also has an  off-diagonal heavy Majorana matrix,
but has $a\ll b,c$ and so leads to the SMA solution. 
\footnote{The latest data from the SNO collaboration \cite{SNO} 
rather strongly disfavours the SMA solution \cite{bah2001}.} 
FC3 is also taken from \cite{SRHND2},
and is an example of a model with an approximately diagonal
heavy Majorana matrix in the $U(1)$ charge basis.
Using the analytic results in \cite{SRHND2} we find
the approximate expectations for the 
experimentally accesible quantities ($\theta_{23}$, $\theta_{13}$, 
$\theta_{12}$ and $R\equiv |\Delta m_{21}^2|/|\Delta m_{32}^2|$)
where $\Delta m^2_{ij} \equiv m^2_i - m^2_j$
as given in Table \ref{FC}. 
Thus FC1 is suitable for the LMA solution, FC2 for the SMA solution,
FC3 for the LMA but with a larger CHOOZ angle than FC1,
and FC4 is a model without SRHND which is consequently expected to give
a larger value of $R$ than models FC1-FC3 which all have SRHND.
\footnote{As a side remark we mention that for neutrinoless double beta 
decay, in flavour models which make use of the seesaw mechanism, one 
never expects that the effective Majorana neutrino mass 
($\langle m_{\nu} \rangle=(m_{LL})_{11}$)
measured in double beta decay is exactly zero. However, these 
models produce a hierarchical spectrum of left-handed neutrinos 
and thus one expects $\langle m_{\nu} \rangle$ to be small. In 
the models we have studied in this paper, one typically gets 
$\langle m_{\nu} \rangle \sim 10^{-3}$ $eV$, albeit depending on 
the model and with a rather larger uncertainty. }

\begin{table}
 \begin{center}
  \begin{tabular}{||l|r|r|r|r|r|r|r|r|r|r|r||} 
\hline
Models & $l_1$ & $l_2$ & $l_3$ & $n_1$ & $n_2$ & $n_3$ & $\sigma$ &
$\theta_{23}$ & $\theta_{13}$ & $\theta_{12}$ & $R$ \cr
\hline
FC1 &  -2 & 0 & 0 & -2 & 1 & 0 & 0 & 1 & $\lambda^2$ & 1 & $\lambda^4$ \cr
\hline
FC2 &  -3 & -1 & -1 & -3 & 0 & -1 & 3 & 1 & 
$\lambda^2$ &$\lambda^2$  & $\lambda^4$ \cr
\hline
FC3 &  -1 & 1 & 1 & $\frac{1}{2}$ & 0 & -$\frac{1}{2}$ & -1 & 
1 & $\lambda$ & 1 & $\lambda^4$ \cr
\hline
FC4 &  -1 & 1 & 1 & $\frac{1}{2}$ & -$\frac{1}{2}$ & -$\frac{1}{2}$ & -1 
& 1 & $\lambda$ & - & - \cr
\hline
\end{tabular}
\end{center}
\caption{Flavour charges (FC) for four models, as discussed
in the text, and approximate expectations for 
$\theta_{23}$, $\theta_{13}$, $\theta_{12}$ and $R$ for the 
four different models.}
\label{FC}
\end{table}

Figure 1 shows the distributions for the solar ($s_{\odot} \equiv
4 \sin\theta_{12}^2(1-\sin\theta_{12}^2)$), atmospheric ($s_{Atm} \equiv 
4 \sin\theta_{23}^2(1-\sin\theta_{23}^2)$) and CHOOZ ($s_{C} \equiv 
4 \sin\theta_{13}^2(1-\sin\theta_{13}^2)$) angles as well as 
for $R\equiv |\Delta m_{21}^2|/|\Delta m_{32}^2|$ 
for the four models given in table 1. As discussed above, the detailed 
shape of the distributions is different to the one we calculated 
previously \cite{discr} using {\it real} coefficients. The positions of 
the peaks of the various distributions, however, did not change allowing for 
complex phases.

\begin{figure*}
\begin{picture}(0,0)
\put(0,-60)
{\mbox{\epsfig{file=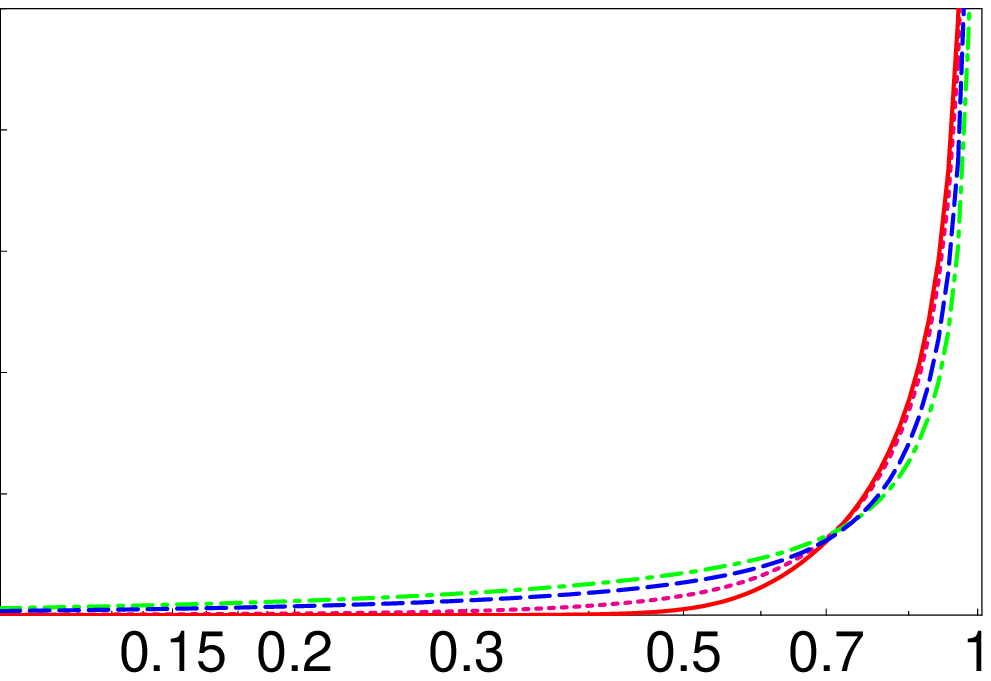,height=6.0cm,width=6.0cm}}}
\end{picture}
\begin{picture}(0,0)
\put(200,-60)
{\mbox{\epsfig{file=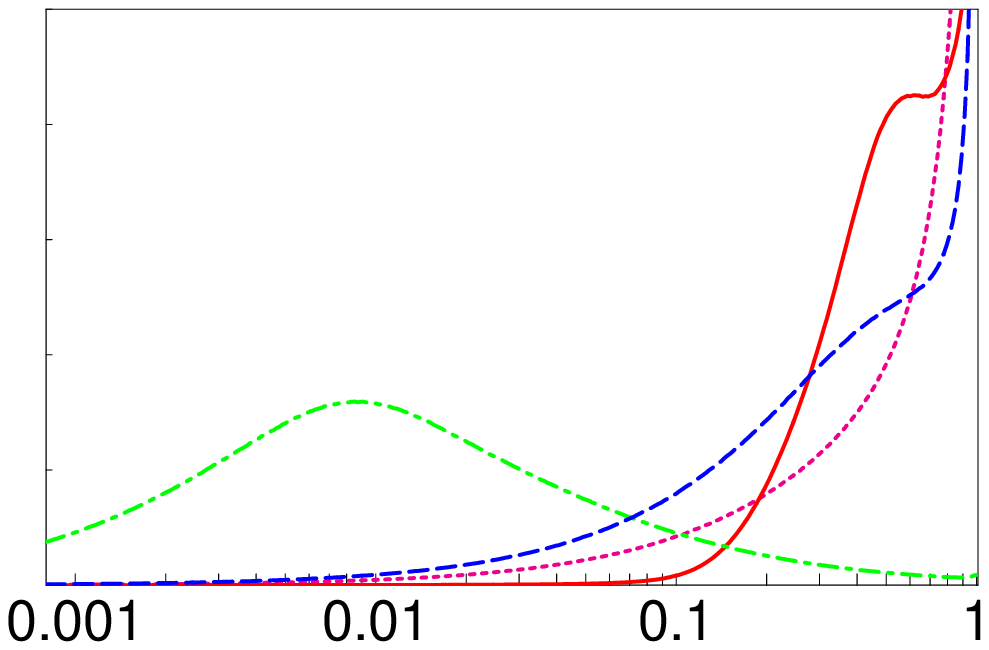,height=6.0cm,width=6.0cm}}}
\end{picture}
{\vskip10mm\hskip50mm {\Large $s_{Atm}$ \hskip60mm $s_{\odot}$}
\vskip0mm}
\begin{picture}(0,55)
\put(0,-95)
{\mbox{\epsfig{file=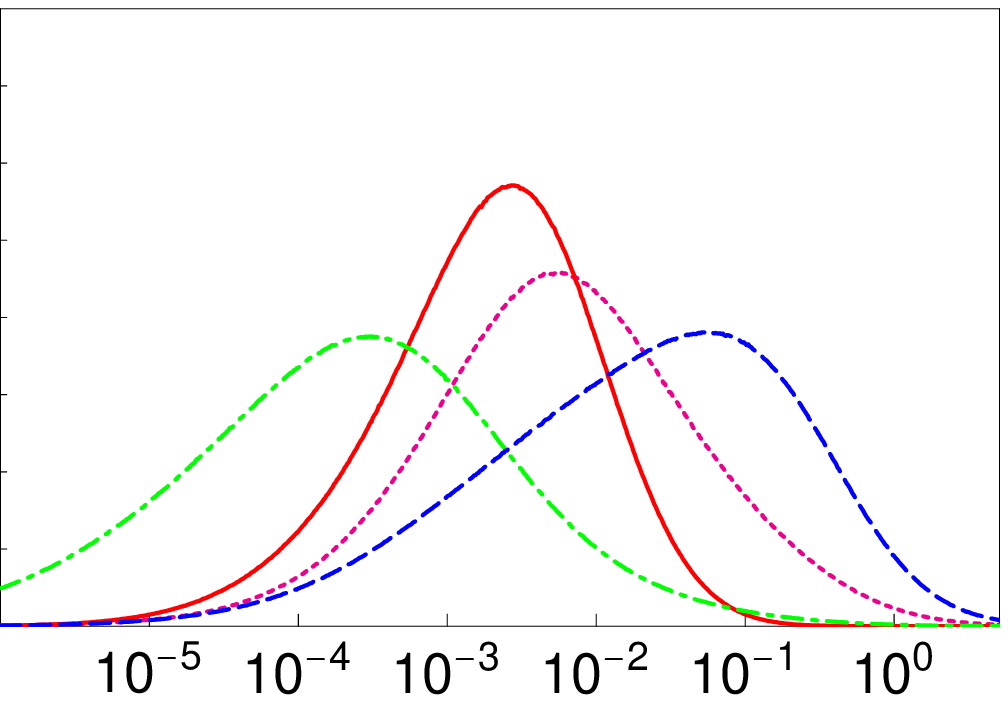,height=6.0cm,width=6.0cm}}}
\end{picture}
\begin{picture}(0,55)
\put(200,-95)
{\mbox{\epsfig{file=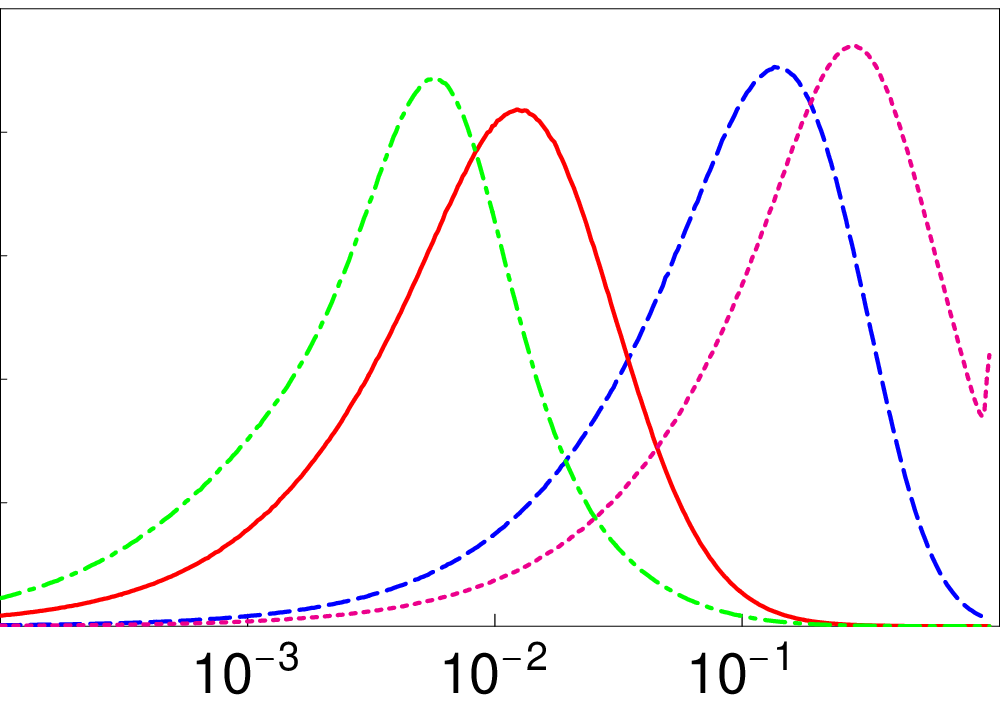,height=6.0cm,width=6.0cm}}}
\end{picture}
{\vskip22mm\hskip50mm  $R$ \hskip70mm {\Large $s_C$}
\vskip2mm}
\caption[allfig]{Theoretical distributions for the predictions of neutrino 
mass and mixing parameters for four selected see-saw models:
FC1 (full), FC2 (dot-dashes), FC3 (thick dots), FC4 (dashes). 
Matrix coefficients are randomly chosen in the interval 
$[\sqrt{2\lambda},1/\sqrt{2\lambda}]\times e^{i\phi}$, with 
$\phi \Rightarrow [0,2\pi]$.
The vertical axis in each panel (deliberately not labelled)
represents the {\em logarithmically binned} 
distributions with correct relative normalisation
for each model, with heights plotted on a linear scale in arbitrary units.}
\label{fig:allplots}
\end{figure*}

\begin{figure}
\setlength{\unitlength}{1mm}
\begin{picture}(0,55)
\put(0,-5)
{\mbox{\epsfig{file=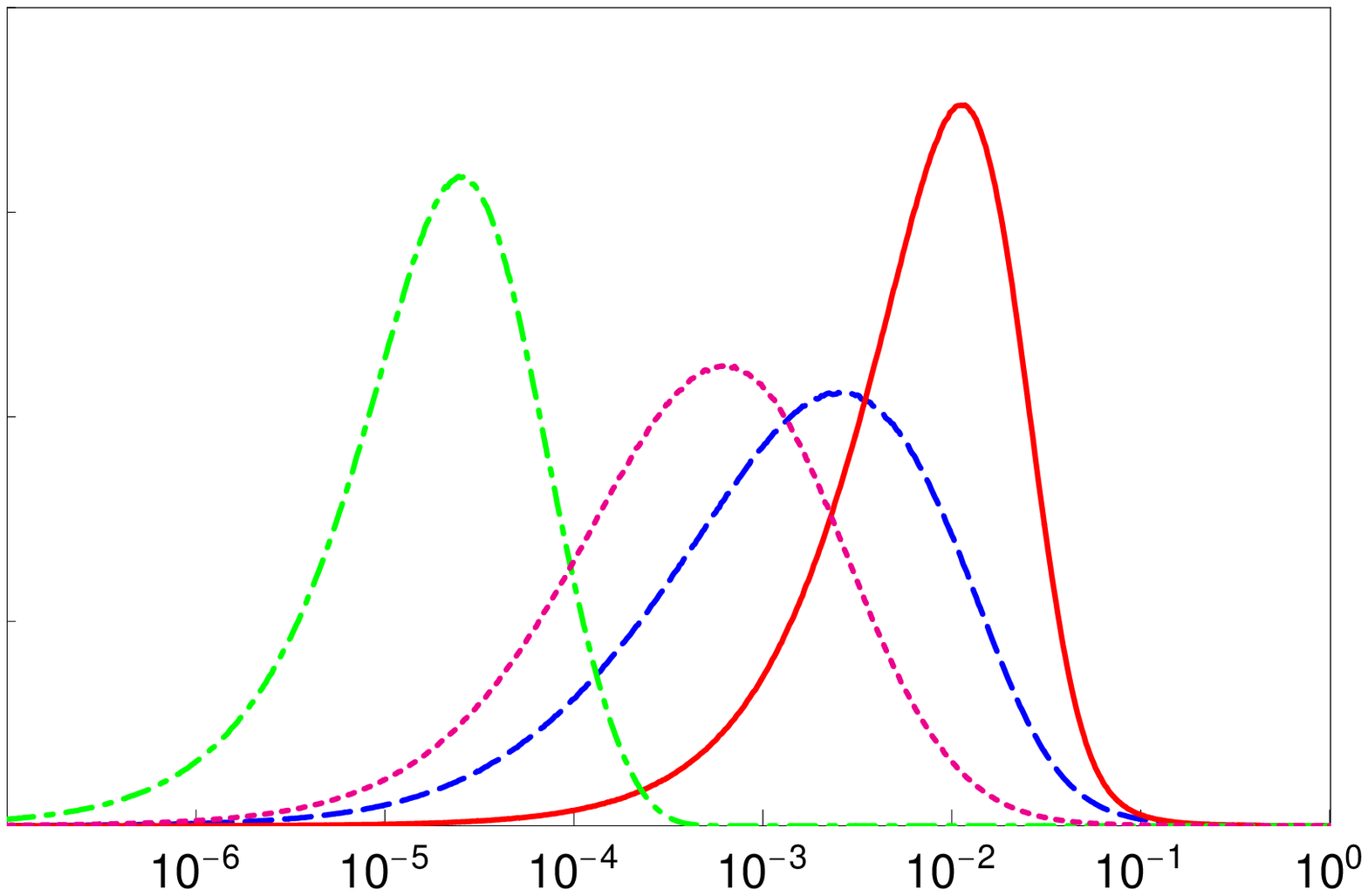,height=8.0cm,width=6.0cm}}}
\end{picture}
{\vskip-16mm\hskip50mm {\Large $\epsilon$}
\vskip-2mm}
\caption[allfig]{Plots of $\epsilon$ for the four different models of 
table 1. Plot style for different models follows fig. \ref{fig:allplots}}
        \label{fig:eps1}
\end{figure}

Figure \ref{fig:eps1} shows the distributions in the leptogenesis 
observable $\epsilon$ for the models FC1-FC4. From the figures one 
might be tempted to think, that different low energy observables lead 
to different values of $\epsilon_1$ and so might be distinguished. 
This is not true, and we now show that any of the models can be 
modified to give any desired value of $\epsilon_1$, while keeping 
the low energy observables approximately unchanged.

Let us consider as an example the model FC3 discussed above,
which predicts the LMA solution and a relatively large CHOOZ angle. 
Model FC3 gives (neglecting the coefficients and assuming 
$\bar{\lambda}= \lambda$) the following Dirac 
and Majorana mass matrices:
\beq
Y_{\nu}^{FC3} = 
\left( \begin{array}{ccc}
 \lambda^{1/2} &  \lambda & \lambda^{3/2} \\ 
 \lambda^{3/2} & \lambda &  \lambda^{1/2} \\ 
 \lambda^{3/2} & \lambda &  \lambda^{1/2} 
\end{array}
\right)
\label{yukfc3}
\end{equation}
\beq
Y_{RR}^{FC3} = 
\left( \begin{array}{ccc}
 1 &  \lambda^{1/2} & \lambda \\ 
 \cdot & \lambda &  \lambda^{3/2} \\ 
 \cdot & \cdot &  \lambda^{2} 
\end{array}
\right)
\label{yukRRfc3}
\end{equation}
which after see-sawing give the following leading order structure 
for $m_{LL}$:
\beq
m_{LL}^{FC3}
\sim 
\left( \begin{array}{ccc}
\lambda & 1 & 1    \\
1 & \lambda^{-1} & \lambda^{-1}    \\
1 & \lambda^{-1} & \lambda^{-1}   
\end{array}
\right)
+ {\cal O}
\left( \begin{array}{ccc}
\lambda & \lambda & \lambda    \\
\lambda & \lambda & \lambda    \\
\lambda & \lambda & \lambda   
\end{array}
\right).
\label{SRHND3}
\eeq
Note, that FC3 has a right-handed neutrino mass matrix which is 
diagonal to leading order and it is the lightest (third) right-handed 
neutrino which gives the dominant contribution to $m_{LL}$.
The estimate for the asymmetry parameter is given in Eq.\ref{a1},
where it is clear that $e \sim \lambda^{1/2} $. In order to change 
$\epsilon_1$ we must reduce $e$. This may be achieved by adjusting
the $l_i$ charges in such a way that the Dirac neutrino
matrix just gets multiplied by an overall scaling factor
compared to eq. \ref{yukfc3}, while the heavy Majorana
Yukawa matrix remains unchanged. The rescaling of the Dirac Yukawa matrix
implies that the coupling $e$ is made smaller, and hence the scale of
right-handed neutrino masses must be reduced in order to maintain
the same value of $m_3$. This will lead to a different value of $\epsilon_1$ 
without changing the other low energy observables at all. 

\begin{figure}
\setlength{\unitlength}{1mm}
\begin{picture}(0,55)
\put(0,-10)
{\mbox{\epsfig{file=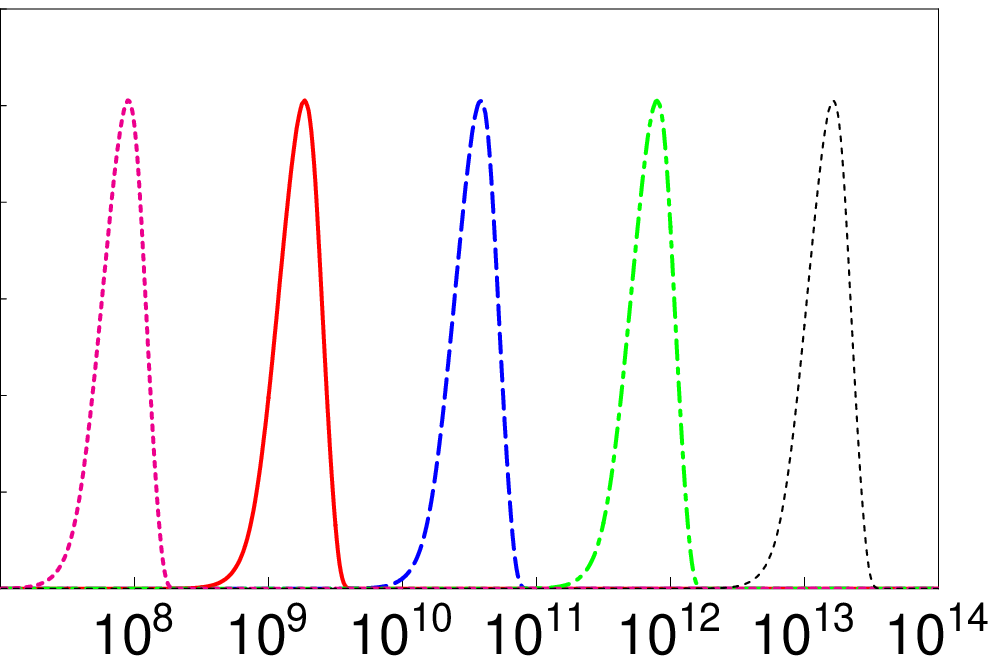,height=6.0cm,width=6.0cm}}}
\end{picture}
\begin{picture}(0,55)
\put(70,-10)
{\mbox{\epsfig{file=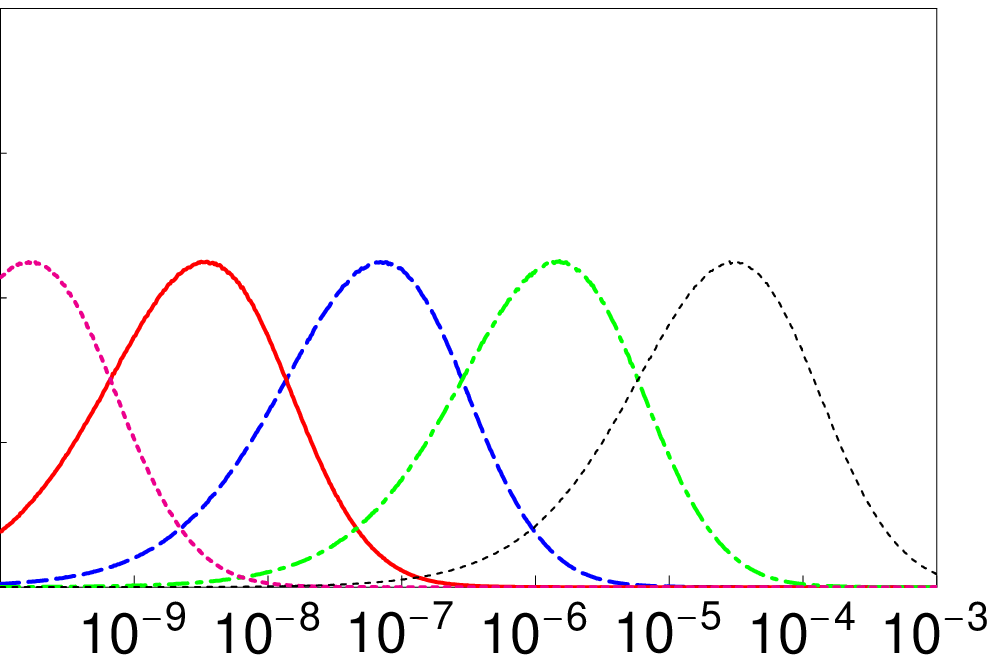,height=6.0cm,width=6.0cm}}}
\end{picture}
{\vskip-1mm\hskip40mm {{$M_1$ [GeV]}\hskip65mm {$\epsilon_1$}}
\vskip2mm}
\caption[allfig]{Plots of the mass of the lightest right-handed 
neutrino $M_1$ (left) and $\epsilon_1$ (right) 
for the five different models of table \ref{FC3}. From right   
to left: FC3a-FC3e.}
        \label{fig:eps8t12}
\end{figure}

This qualitative conclusion is supported by our numerical results.
In table \ref{FC3} we give sets of charges for variants of the model FC3 
of table \ref{FC}, which lead to a simple rescaling of $Y_{\nu}$,
\beq
Y_{\nu}^{FC3(a,b,c,d,e)}=(\lambda^{(1,2,3,4,5)})Y_{\nu}^{FC3}
\eeq
and hence the scale of right-handed neutrino masses
as shown in fig \ref{fig:eps8t12}.

All of these models were constructed to preserve the 
low-energy phenomenology, and in fact we have checked that they lead to 
identical predictions for $s_{Atm}$, $s_{\odot}$, $s_C$ and $R$ as FC3. 
Figure \ref{fig:eps8t12} shows the resulting values of $M_1$ and $\epsilon$. 
Note that the lightest right-handed neutrino mass for FC3a, FC3b and FC3c 
in Fig.\ref{fig:eps8t12} is above the reheat temperature allowed by the 
gravitino constraint \cite{gravitino}. This figure explicitly demonstrates 
that it is possible to completely decouple the predictions for leptogenesis 
from low energy observables.

\begin{table}
 \begin{center}
  \begin{tabular}{||l|r|r|r|r|r|r|r||} 
\hline
Models & $l_1$ & $l_2$ & $l_3$ & $n_1$ & $n_2$ & $n_3$ & $\sigma$ \cr
\hline
FC3a &  -2 & 2 & 2 & $\frac{1}{2}$ & 0 & -$\frac{1}{2}$ & -1 \cr
& & & & & & & \cr
\hline
FC3b &  -3 & 3 & 3 & $\frac{1}{2}$ & 0 & -$\frac{1}{2}$ & -1 \cr
& & & & & & & \cr
\hline
FC3c &  -4 & 4 & 4 & $\frac{1}{2}$ & 0 & -$\frac{1}{2}$ & -1 \cr
& & & & & & & \cr
\hline
FC3d &  -5 & 5 & 5 & $\frac{1}{2}$ & 0 & -$\frac{1}{2}$ & -1 \cr
& & & & & & & \cr
\hline
FC3e &  -6 & 6 & 6 & $\frac{1}{2}$ & 0 & -$\frac{1}{2}$ & -1 \cr
& & & & & & & \cr
\hline
\end{tabular}
\end{center}
\caption{``Variants'' of the flavour model FC3 of table \ref{FC}. All 
of these models give exactly the same distributions for the low energy 
neutrino observables. They differ, however, in their predicted values 
for the leptogenesis obervable $\epsilon$.}  
\label{FC3}
\end{table}

How well do the analytic estimates for $\epsilon$ discussed previously 
agree with the numerical results? In terms of our small expansion 
parameter $\lambda \simeq 0.22$ and inserting the flavour charges 
for the models FC3 (FC3a, FC3b, FC3c, FC3d and FC3e) into eq. 
\ref{a} one finds: 
\begin{equation}
\epsilon_1^{(a)} \simeq \frac{3}{32 \pi} \lambda^3 
\hskip2mm (\lambda^5,\lambda^7,\lambda^9,\lambda^{11},\lambda^{13})
\label{estimate1}
\end{equation}
numerically $3\times 10^{-4}$ ($2 \times 10^{-5}$, $7 \times 10^{-7}$,  
$4\times 10^{-8}$, $2 \times 10^{-9}$ and $8\times 10^{-11}$) which 
coincides approximately with the peaks of the 
distributions in $\epsilon$ shown in fig. \ref{fig:eps8t12}.
Recall that model FC3 predicts a right-handed neutrino mass matrix 
with the dominant neutrino being the lightest one (case a, discussed 
in section 3.2). 

Model FC3 produces predictions for low-energy neutrino phenomenology 
consistent with the large angle MSW solution of the solar neutrino 
problem. It is interesting to ask whether this solution is the only 
one for which one can decouple $\epsilon_1$ from the low-energy 
observables.

\begin{figure}
\setlength{\unitlength}{1mm}
\begin{picture}(0,60)
\put(-10,-125)
{\mbox{\epsfig{file=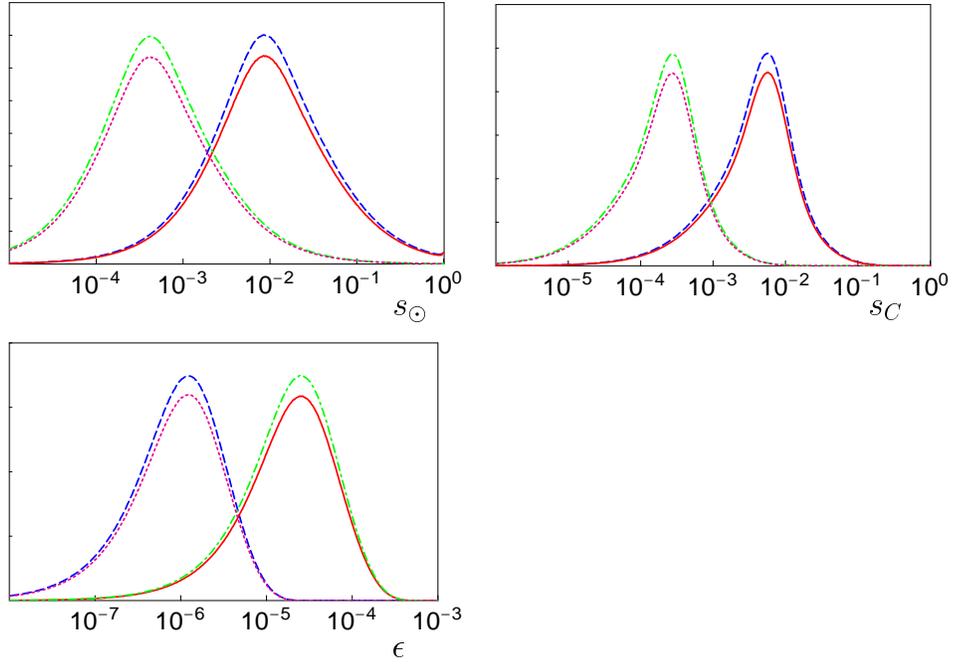,height=22.0cm,width=16.0cm}}}
\end{picture}
\vskip20mm
\caption[allfig]{Plots of $s_{\odot}$ (top left), $s_C$ (top right), 
$\epsilon$ (bottom left) and $Y_B$ (bottom right) for the four different 
models of table \ref{FC2}. The full line is FC2, the dashed line FC2a, 
the dotted line FC2b and the dash-dotted line FC2c. Note, that the 
distributions for FC2a and FC2b have been multiplied by a factor of 
$1.1$, see text. }
        \label{fig:eps15t18}
\end{figure}

In order to investigate this problem we have constructed variants of FC2, 
predicting a small angle MSW solution to the solar neutrino problem. The 
corresponding charges are given in Table. \ref{FC2}. All models in this 
table produce exactly the same distributions in $R$ and $s_{Atm}$ as model 
FC2, but lead to different predictions for $s_{\odot}$, $s_C$ and  
$\epsilon_1$ as is demonstrated in Fig. \ref{fig:eps15t18}. Note that 
we have multiplied the distributions for FC2a and FC2b by a factor of 
$1.1$, since otherwise the curves would completely overlap in some of 
the variables.

\begin{table}
 \begin{center}
  \begin{tabular}{||l|r|r|r|r|r|r|r||} 
\hline
Models & $l_1$ & $l_2$ & $l_3$ & $n_1$ & $n_2$ & $n_3$ & $\sigma$ \cr
\hline
FC2 &  -3 & -1 & -1 & -3 & 0 & -1 & 3 \cr
\hline
FC2a & -4 & -2 & -2 & -3 & 0 & -1 & 3 \cr
\hline
FC2b & -4 & -1 & -1 & -3 & 0 & -1 & 3 \cr
\hline
FC2c & -5 & -2 & -2 & -3 & 0 & -1 & 3 \cr
\hline
\end{tabular}
\end{center}
\caption{``Variants'' of the flavour model FC2 of table \ref{FC}. 
FC2 and FC2a give the same distributions for the low energy observables, 
with an expectation for the solar and CHOOZ angle of order $\lambda^2$, 
FC2b and FC2c, on the other hand, lead to an expectation for 
solar and CHOOZ angle of order $\lambda^4$. See fig. \ref{fig:eps15t18}}
\label{FC2}
\end{table}

As can be seen from Fig. \ref{fig:eps15t18} models FC2 and FC2a give the 
same predictions for $s_{\odot}$ and $s_C$, but differ in their predictions 
for $\epsilon_1$. FC2b and FC2c, on the other hand, give  
expectations for $s_{\odot}$ and $s_C$ which are smaller than the one for 
FC2 by about $1.5$ orders of magnitude. Nevertheless, FC2b yields values 
of $\epsilon$ which are very similiar to those of FC2.  
Also FC2c and FC2a have very similar expectations for leptogenesis 
while differing in $s_{\odot}$ and $s_C$. 

It is obviously easy to find models differing in their predictions 
for leptogenesis and at the same time being consistent with SMA MSW. 
Moreover, neither the size of the solar nor the size of the CHOOZ 
angles tell us anything about whether leptogenesis is possible or 
not.

\begin{figure}
\setlength{\unitlength}{1mm}
\begin{picture}(0,55)
\put(0,0)
{\mbox{\epsfig{file=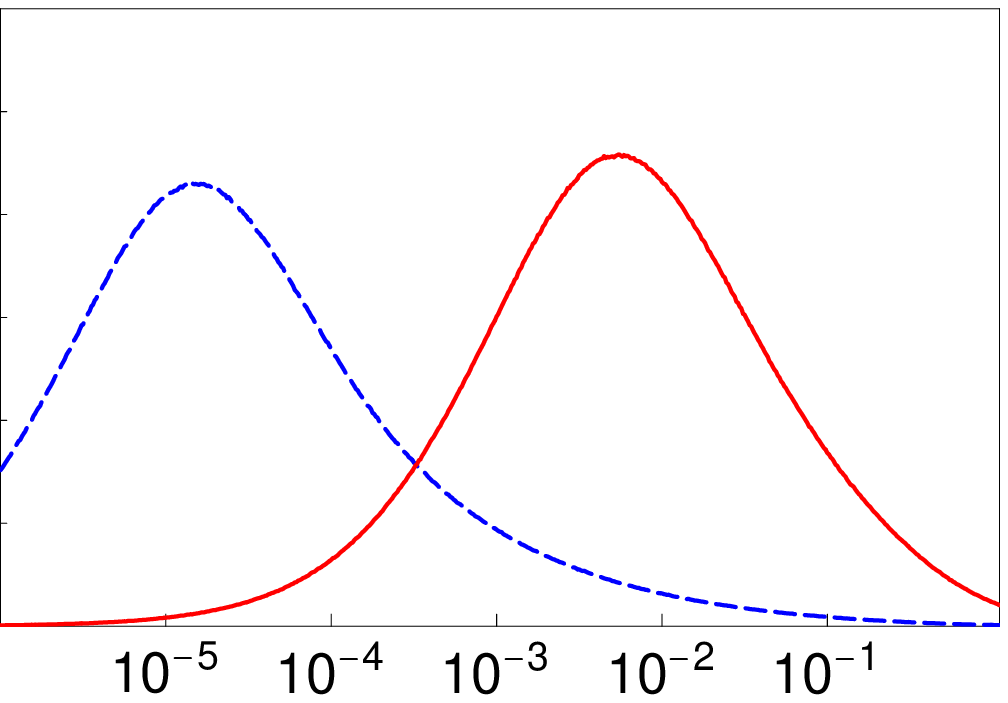,height=6.0cm,width=6.0cm}}}
\end{picture}
\begin{picture}(0,55)
\put(70,0)
{\mbox{\epsfig{file=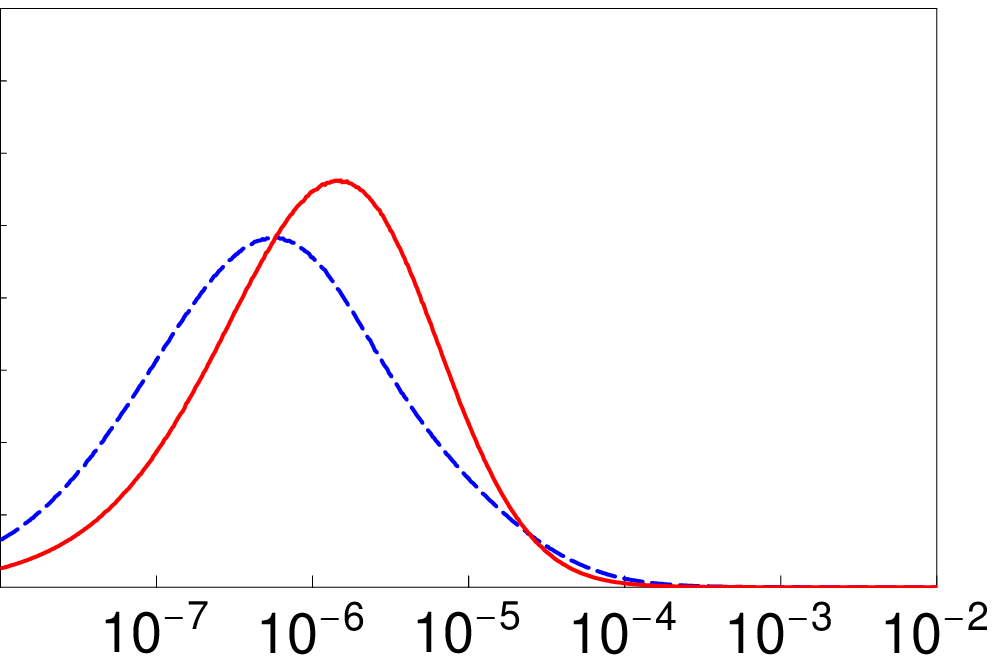,height=6.0cm,width=6.0cm}}}
\end{picture}
{\vskip-10mm\hskip50mm $R$ \hskip70mm {\Large $\epsilon$}
\vskip-2mm}
\caption[allfig]{Plots of $R$ (to the left) and $\epsilon$ (to the 
right) for the 2 different models FC3b (full line) and FC5 (dashed 
line). } 
        \label{fig:fc5}
\end{figure}

Finally we have investigated the question whether a special value of 
$R$ determines $\epsilon_1$. All the models discussed so far 
prefer values of $R > 10^{-4}$. The following assignment of charges 
defines a model (FC5), which prefers larger hierarchies, see fig. 
\ref{fig:fc5},

\begin{equation}
(l_1,l_2,l_3,n_1,n_2,n_3,\sigma) = (3,-3,-3,0,-1/2,1,1),
\end{equation}
while still keeping the atmospheric and solar angles large (and $s_C \ll 1$). 
FC5 therefore is consistent with the LOW solution of the solar neutrino 
problem. Nevertheless, as fig. \ref{fig:fc5} demonstrates FC5 leads to a 
very similar expectation for $\epsilon$ as the model FC3b discussed 
above, which prefers $R$ in the range $R \sim 10^{-3}-10^{-2}$. 

Obviously, any value of $R$ can produce approximately the same 
order-of-magnitude values of $\epsilon_1$. 

As a summary it can be stated that there is a decoupling
between low energy neutrino observables and leptogenesis. We have 
demonstrated this point by constructing a number of different flavour 
models, which give the same predictions for neutrino masses and mixings 
while differing by huge factors in their expectations for leptogenesis. 

On the other hand we have seen that leptogenesis is in principle able to 
resolve the ambiguity between different models which would lead to the 
same low energy neutrino observables, and which otherwise would be 
indistinguishable. Therefore leptogenesis is able to provide information 
about the high energy theory which could not be obtained by low energy 
measurements.

\section{From $\epsilon_1$ to $Y_B$ when the dominant right-handed neutrino
is the lightest}

While the calculation of $\epsilon_1$ is straightforward, once a 
particular model has been specified, the calculation of $Y_B$ 
depends crucially on a number of assumptions made about the 
early universe. In the following calculation we will assume a 
standard hot big bang scenario in which the maximum temperature 
is higher than the largest right-handed neutrino mass we consider, 
such that the right-handed neutrinos can be thermally produced. 
This assumption is necessary if one wants to employ one of the 
parameterizations to the full solution of the Boltzman equations, 
see eqs \ref{dtak1}-\ref{dtak2} and \ref{dnew1}-\ref{dnew3}, discussed 
in section 2.3.

Obviously the following discussion will not be valid, if the universe 
underwent a period of inflation with a rather low reheat temperature, 
as required by the gravitino problem. 

Let us first discuss the different values for $Y_B$ one obtains using 
either eqs \ref{dtak1}-\ref{dtak2} or our parameterization eqs 
\ref{dnew1}-\ref{dnew3}. As an example we will concentrate on the 
variants of the model FC3, discussed in the last section.

\begin{figure}
\setlength{\unitlength}{1mm}
\begin{picture}(0,55)
\put(0,0)
{\mbox{\epsfig{file=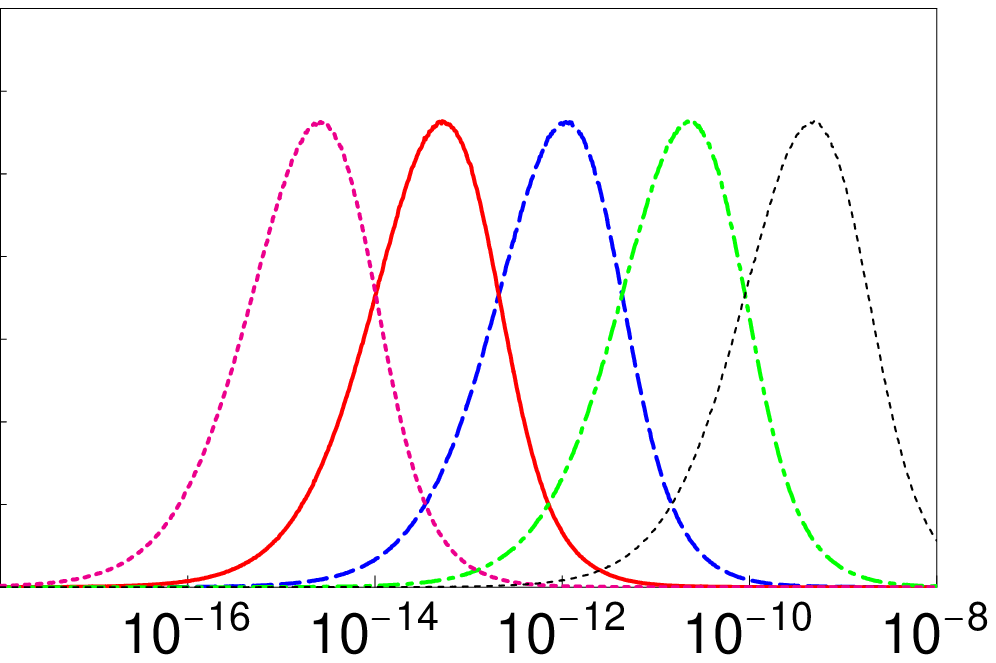,height=6.0cm,width=6.0cm}}}
\end{picture}
\begin{picture}(0,55)
\put(70,0)
{\mbox{\epsfig{file=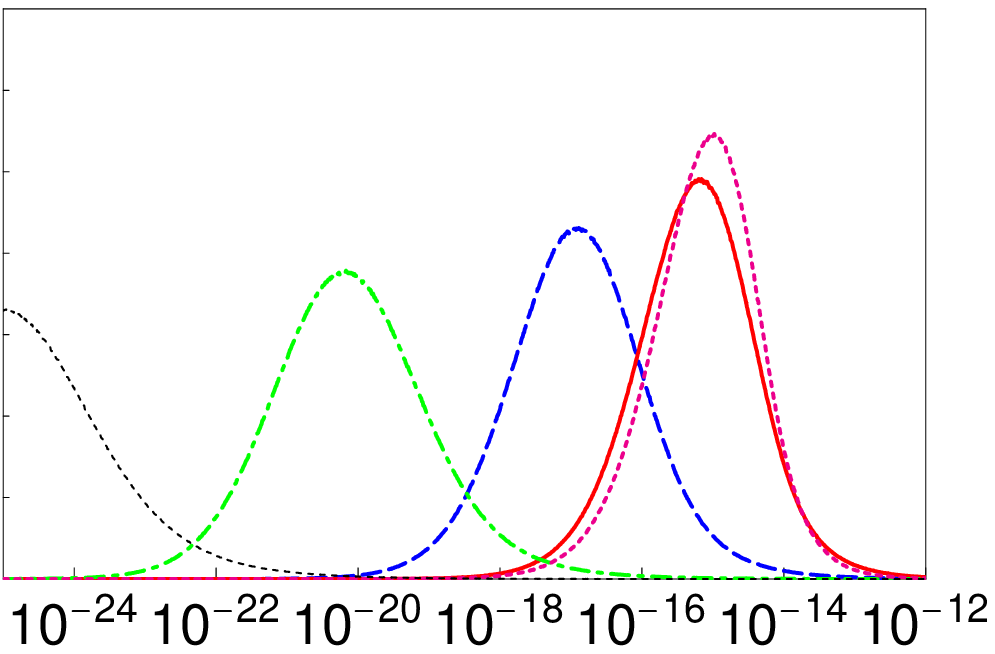,height=6.0cm,width=6.0cm}}}
\end{picture}
{\vskip-10mm\hskip50mm $Y_B$ \hskip70mm $Y_B$
\vskip-2mm}
\caption[allfig]{Plots of $Y_B$ following eqs \ref{dtak1}-\ref{dtak2} 
(to the left) or according to eqs \ref{dnew1}-\ref{dnew3} (to the 
right) for the different variants of model FC3. Line style as in 
fig. \ref{fig:eps8t12}.}
        \label{fig:OldandNew}
\end{figure}

Fig. \ref{fig:OldandNew} shows calculated values of $Y_B$ using the 
two different approximations. Obviously for large values of $M_1$ 
the two different calculations differ by many orders of magnitude. 
Using the simplest approximation, eqs \ref{dtak1}-\ref{dtak2}, it 
seems that larger values of $M_1$ lead to larger values of $Y_B$ 
and that, in principle, one can get $Y_B$ as large as one desires. 
One can trace back this scaling to eq. \ref{a} in section 3.2 and 
to the fact that eqs \ref{dtak1}-\ref{dtak2} do not depend on the 
value of $M_1$. 

On the other hand, using eqs \ref{dnew1}-\ref{dnew3}, which take into 
account the suppresion of $Y_B$ for large values of $M_1$ and 
${\tilde m}_1$ one gets a completely different picture. For large 
values of $M_1$, $Y_B$ is suppressed to negligible values and going 
to smaller values of $M_1$ increases $Y_B$. Note, however, that for 
the smallest values of $M_1$ of the order of ${\cal O}(10^8)$ [GeV] 
$Y_B$ stops growing and never reaches the experimentally preferred 
range of $Y_B \sim (0.5-1)\times 10^{-10}$. All these variants of FC3 
therefore fail the leptogenesis test. 

Since the simple approximation, eqs \ref{dtak1}-\ref{dtak2} 
\cite{yasutaka}, employed similarly by a number of authors \cite{kolb}, 
fails to take into account any $M_1$ dependence of the dilution function 
its use would lead to the opposite conclusion. A careful treatment of 
$d$ seems to be absolutely necessary for a reliable calculation of 
$Y_B$ and we stress again that also our treatment is still only
approximate.

\begin{figure}
\setlength{\unitlength}{1mm}
\begin{picture}(0,80)
\put(8,0)
{\mbox{\epsfig{file=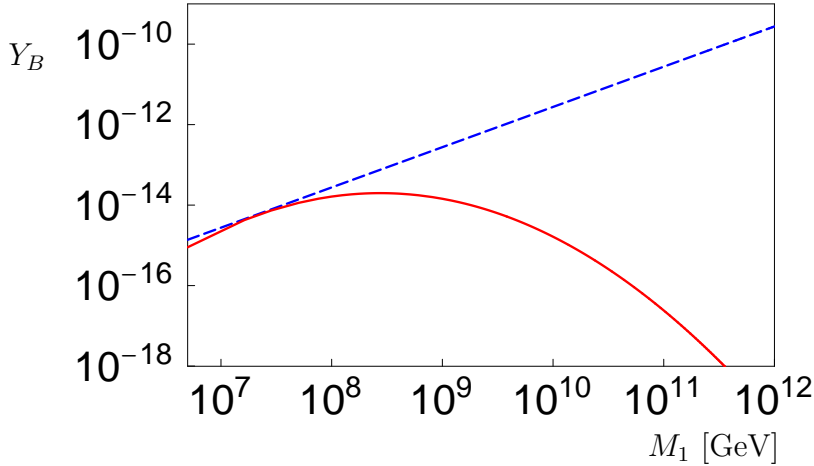,height=10.0cm,width=10.0cm}}}
\end{picture}
{\vskip-75mm\hskip0mm $Y_B$ 
\vskip47mm \hskip85mm $M_1$ [GeV]
\vskip-2mm}
\caption[allfig]{Plots of $Y_B$ following eqs \ref{dtak1}-\ref{dtak2} 
(broken line) or according to eqs \ref{dnew1}-\ref{dnew3} (full line) 
assuming ${\tilde m}_1 = 0.05$ eV. For this plot we have assumed that 
$\epsilon_1$ scales as given by eq. \ref{a1}.}
        \label{fig:CompFit}
\end{figure}

In fig. \ref{fig:CompFit} we plot $Y_B$ as defined in eqs \ref{defyl} 
and \ref{defyb} with $\epsilon_1$ estimated from eq. \ref{a1} and with 
$d$ calculated by a) the simple approximation, defined in eqs 
\ref{dtak1}-\ref{dtak2} and b) our fit to the exact solution of the 
Boltzman equations \cite{FitExact}, defined in eqs \ref{dnew1}-\ref{dnew3}. 
For both calculations we fixed ${\tilde m}_1$ to ${\tilde m}_1 = 0.05$ eV. 
For small values of $M_1$ both approximations agree quite well, whereas 
for $M_1$ larger than $M_1 \sim 10^9$ GeV the expectations from the 
different approximations differ by many orders of magnitude. 

One can understand the failure of the models FC3 to produce the 
correct value of $Y_B$ on the basis of the discussion presented 
in section 2.3. \footnote{For the analytic estimations of 
$\epsilon_1$ and ${\tilde m}_1$ we assumed that the right-handed 
neutrino mass matrix is diagonal, which is approximately true 
for the variants of FC3.}
From the analytic estimates presented in section 2.3 one finds that 
in models where the dominant right-handed neutrino is the lightest, 
${\tilde m}_1$ depends on the same combination of Yukawas as the value 
of the heaviest neutrino mass, fixed by the atmospheric neutrino mass 
scale. Thus, for these models ${\tilde m}_1 \sim m_{3} \sim 0.05$ eV. 
At such large values of ${\tilde m}_1$, however, the dilution function 
$d$, see Fig. \ref{fig:pluemi}, is heavily suppressed for values of 
$M_1$ larger than $M_1 \sim 10^8$ GeV. Thus, although larger values of 
$M_1$ lead to larger values of $\epsilon_1$, see eq. \ref{a1}, the price 
one hast to pay for such large masses in the dilution function always 
overcompensates and $Y_B$ in these models can never be larger than 
$Y_B \sim 10^{-14}$ as is demonstrated in fig. \ref{fig:CompFit}.

Since all models where the dominant right-handed neutrino is the lightest 
share the feature ${\tilde m}_1 \sim m_{3}$, we conclude that upon 
use of eqs \ref{dnew1}-\ref{dnew3} they all fail the leptogenesis 
test. In the next section we will 
therefore study models in which the dominant 
right-handed neutrino is the heaviest.

\section{Leptogenesis prefers models where the dominant right-handed neutrino
is the heaviest}

In the previous section we have seen that although leptogenesis
is decoupled from the low energy neutrino observables,
nevertheless leptogenesis is capable of resolving
the ambiguities between classes of models which would
otherwise lead to the same experimental predictions.
As an example of the power of leptogenesis to give
information about the high energy theory,
in this section we show that leptogenesis prefers
models where the dominant right-handed neutrino is the
heaviest one and discuss the implications of this for unified models.
According to our analytic estimates
we expect this class of models to yield a lightest right-handed
neutrino mass which is lighter than in the previous case, and
hence more acceptable from the point of view of the gravitino
constraint. In addition these models may be more consistent
with GUTs.

As a first example of a case (b) model we consider
the charge vector 
\beq
(l_1,l_2,l_3,n_1,n_2,n_3,\sigma)=(-3,1,1,9,1,-1,2),
\label{caseb}
\eeq
which defines a new model called FC9. The charges in Eq.\ref{caseb} lead to
\begin{equation}
Y_{\nu}^{FC9}  
\sim 
\left( \begin{array}{ccc}
\lambda^{6} & \lambda^{2} & \lambda^{4}    \\
\lambda^{10} & \lambda^{2} & 1   \\
\lambda^{10} & \lambda^{2} & 1   
\end{array}
\right)
\label{yukb}
\end{equation}
and an approximately diagonal Majorana matrix
\begin{equation}
Y_{RR}^{FC9}  
\sim 
\left( \begin{array}{ccc}
\bar{\lambda}^{20} & \bar{\lambda}^{12} & \bar{\lambda}^{10}    \\
\bar{\lambda}^{12} & \bar{\lambda}^4 & \bar{\lambda}^2   \\
\bar{\lambda}^{10} & \bar{\lambda}^2 & 1   
\end{array}
\right)
\label{mrrb}
\end{equation}
Now if we take $\bar{\lambda}= \sqrt{\lambda}$,
this leads to the contributions from the heaviest (dominant),
intermediate, and lightest right-handed neutrino, respectively,
to the effective Majorana matrix of the order of,
\beq
m_{LL}^{FC9}
\sim 
\left( \begin{array}{ccc}
\lambda^8 & \lambda^4 & \lambda^4    \\
\lambda^4 & 1 & 1    \\
\lambda^4 & 1 & 1  
\end{array}
\right)
+ {\cal O}
\left( \begin{array}{ccc}
\lambda^2 & \lambda^2 & \lambda^2    \\
\lambda^2 & \lambda^2 & \lambda^2    \\
\lambda^2 & \lambda^2 & \lambda^2   
\end{array}
\right)
+ {\cal O}
\left( \begin{array}{ccc}
\lambda^2 & \lambda^6 & \lambda^6    \\
\lambda^6 & \lambda^{10} & \lambda^{10}    \\
\lambda^6 & \lambda^{10} & \lambda^{10}   
\end{array}
\right).
\label{SRHND4}
\eeq
By inspection we see that the model predicts $\theta_{12}\sim 1$,
$\theta_{13}\sim \lambda^4$, and, from the order $\lambda^2$
accuracy of the SRHND condition, $R\sim \lambda^4$. 
It may therefore be suitable for one of
the large mixing angle solar solutions, either LMA or LOW.
Assuming $\bar{\lambda}= \sqrt{\lambda}$, the lightest right-handed
neutrino mass is predicted to be $X'\sim \lambda^{10}Y$,
or  $X'\sim 3.10^{-7}Y\sim 10^{8}$ GeV, which is rather small.
In order to increase $X'$ we need to increase $\bar{\lambda}$. 

\begin{figure}
\setlength{\unitlength}{1mm}
\begin{picture}(0,55)
\put(0,0)
{\mbox{\epsfig{file=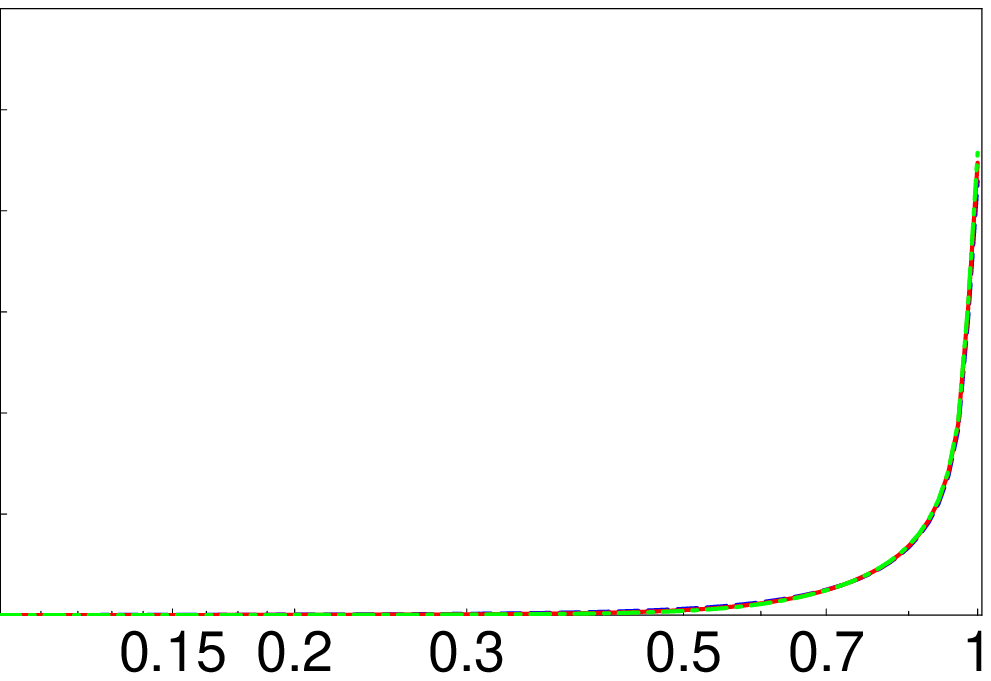,height=6.0cm,width=6.0cm}}}
\end{picture}
\begin{picture}(0,55)
\put(70,0)
{\mbox{\epsfig{file=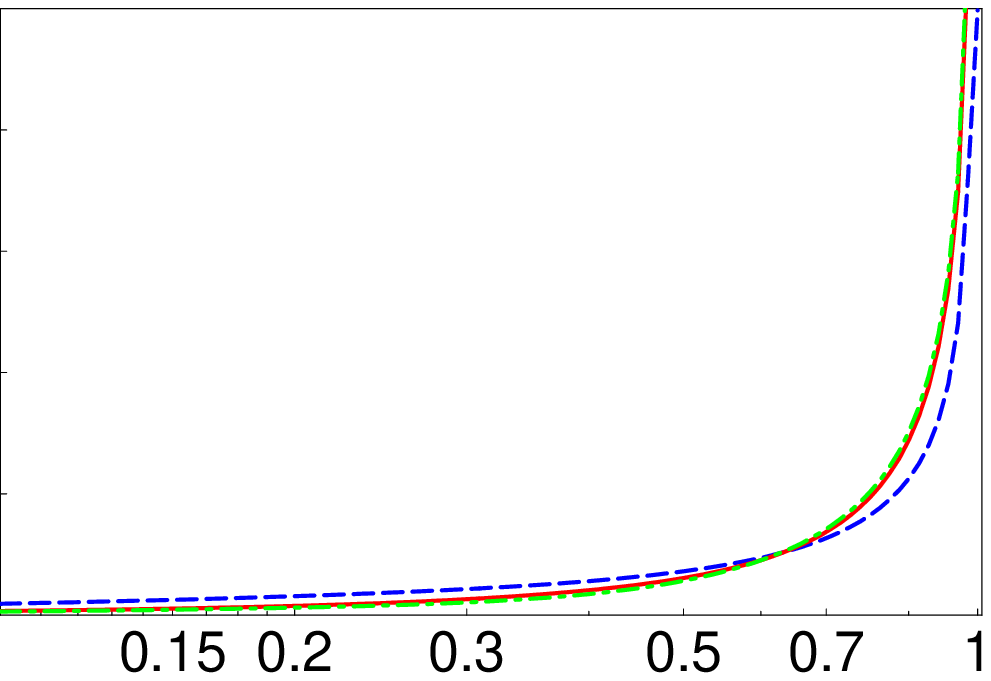,height=6.0cm,width=6.0cm}}}
\end{picture}
{\vskip-10mm\hskip50mm {\Large $s_{Atm}$ \hskip60mm $s_{\odot}$}
\vskip2mm}
\begin{picture}(0,55)
\put(0,0)
{\mbox{\epsfig{file=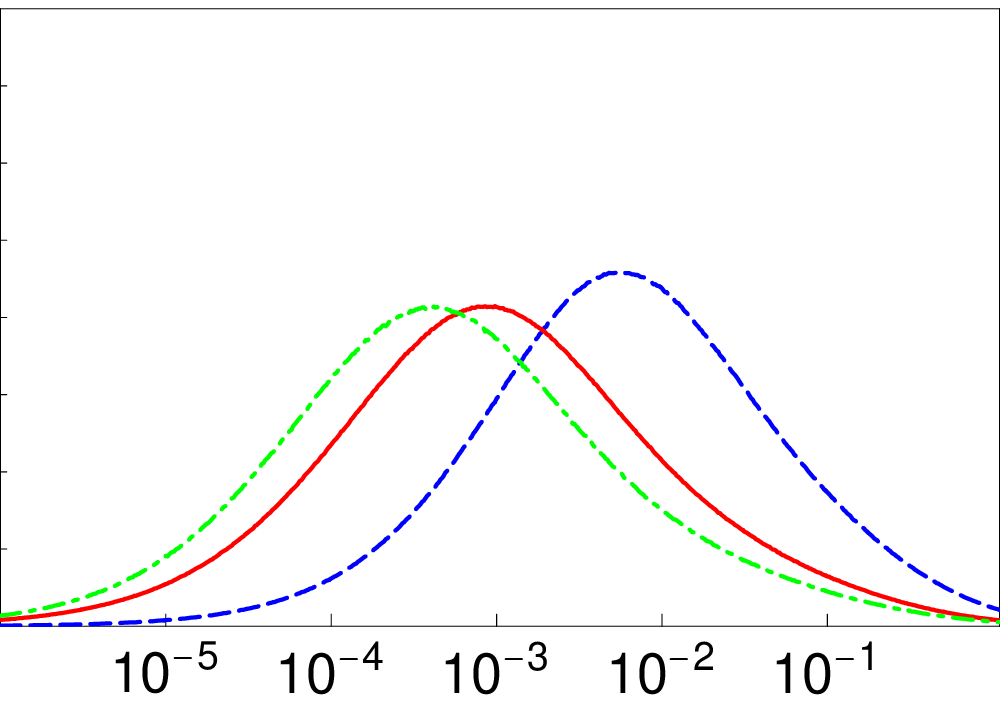,height=6.0cm,width=6.0cm}}}
\end{picture}
\begin{picture}(0,55)
\put(70,0)
{\mbox{\epsfig{file=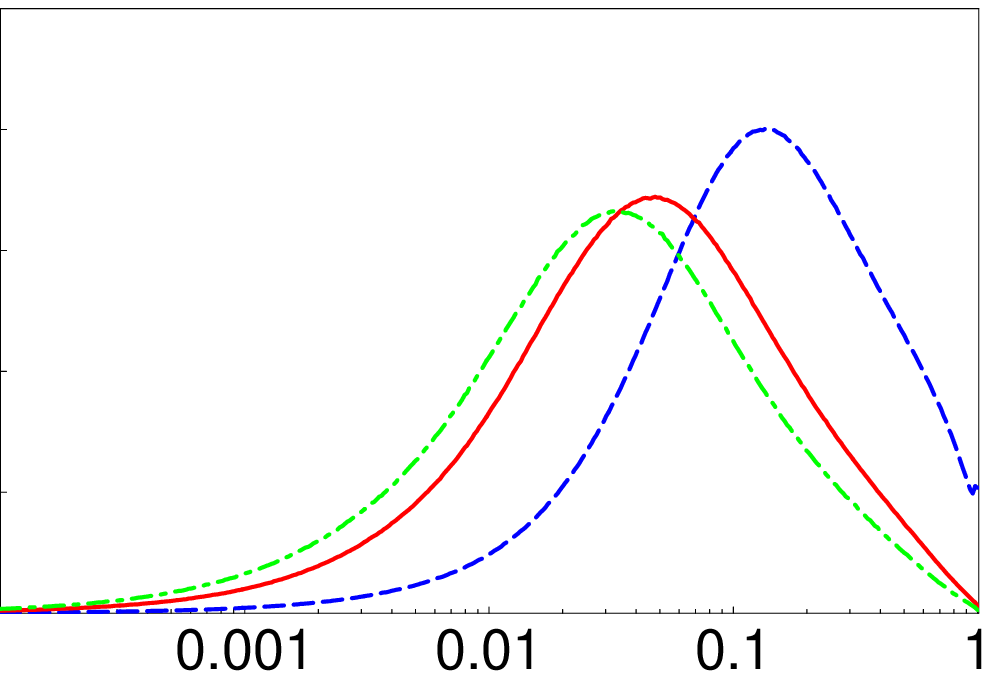,height=6.0cm,width=6.0cm}}}
\end{picture}
{\vskip-10mm\hskip50mm  $R$ \hskip70mm {\Large $s_C$}
\vskip2mm}
\begin{picture}(0,55)
\put(0,0)
{\mbox{\epsfig{file=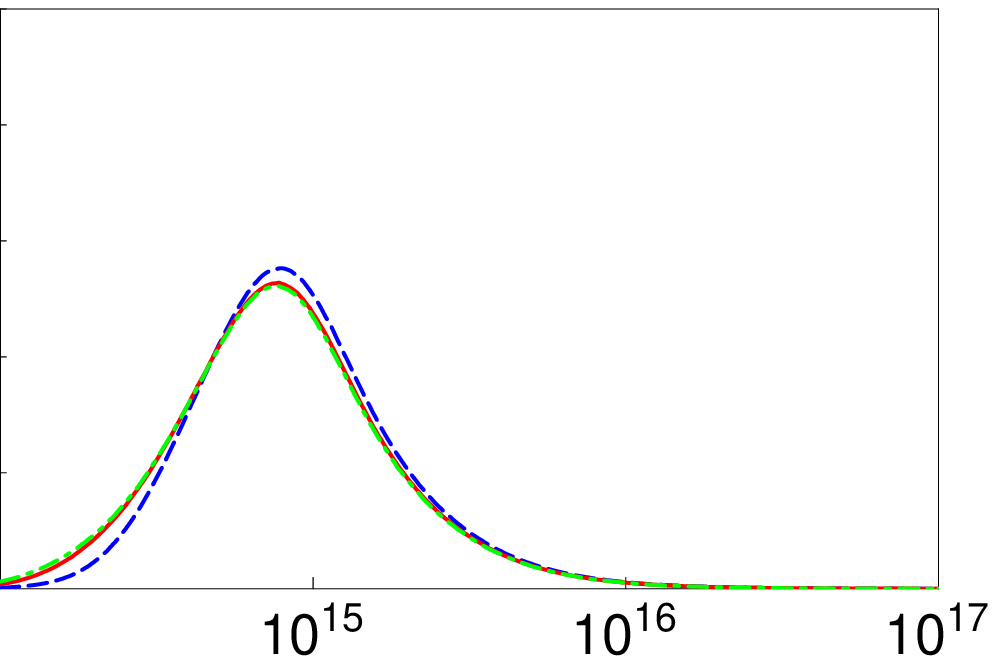,height=6.0cm,width=6.0cm}}}
\end{picture}
\begin{picture}(0,55)
\put(70,0)
{\mbox{\epsfig{file=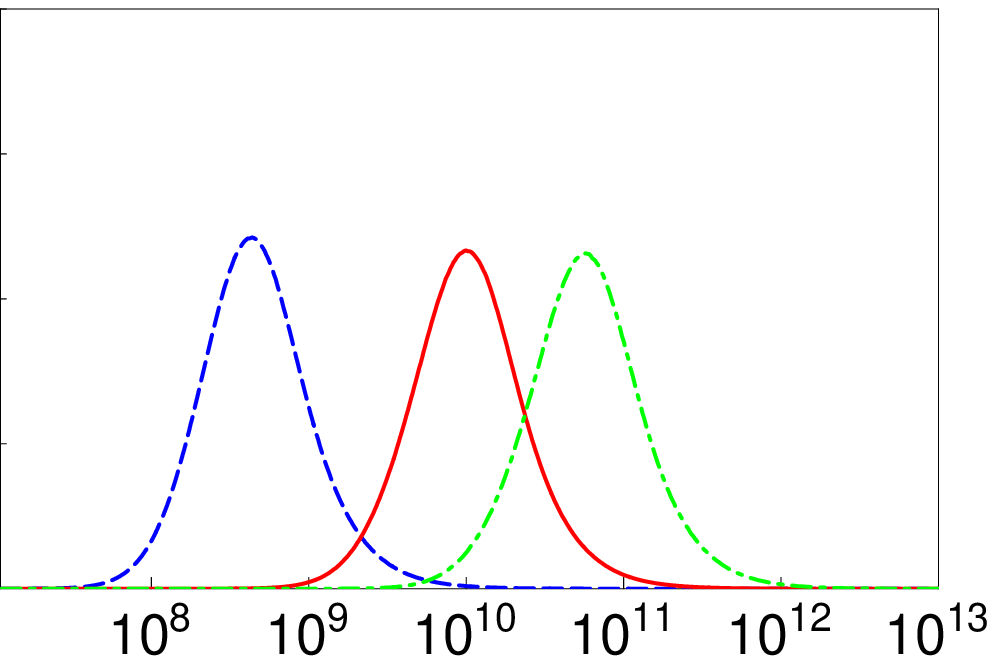,height=6.0cm,width=6.0cm}}}
\end{picture}
{\vskip-10mm\hskip40mm {$\langle \Sigma \rangle$ [GeV]\hskip55mm {$M_1$ [GeV]}}
\vskip2mm}
\caption[allfig]{Plots of (from top left to bottom right: $s_{Atm}$, 
$s_{\odot}$, $R$, $s_C$, $\langle \Sigma \rangle$ and $M_1$ [GeV] 
for the model defined in eq. \ref{caseb}. Dashed line: 
${\bar \lambda}=\sqrt{\lambda}$, full line: ${\bar \lambda}=0.55$, 
dash-dotted: ${\bar \lambda}=0.60$.}
        \label{fig:modb}
\end{figure}

\begin{figure}
\setlength{\unitlength}{1mm}
\begin{picture}(0,55)
\put(0,0)
{\mbox{\epsfig{file=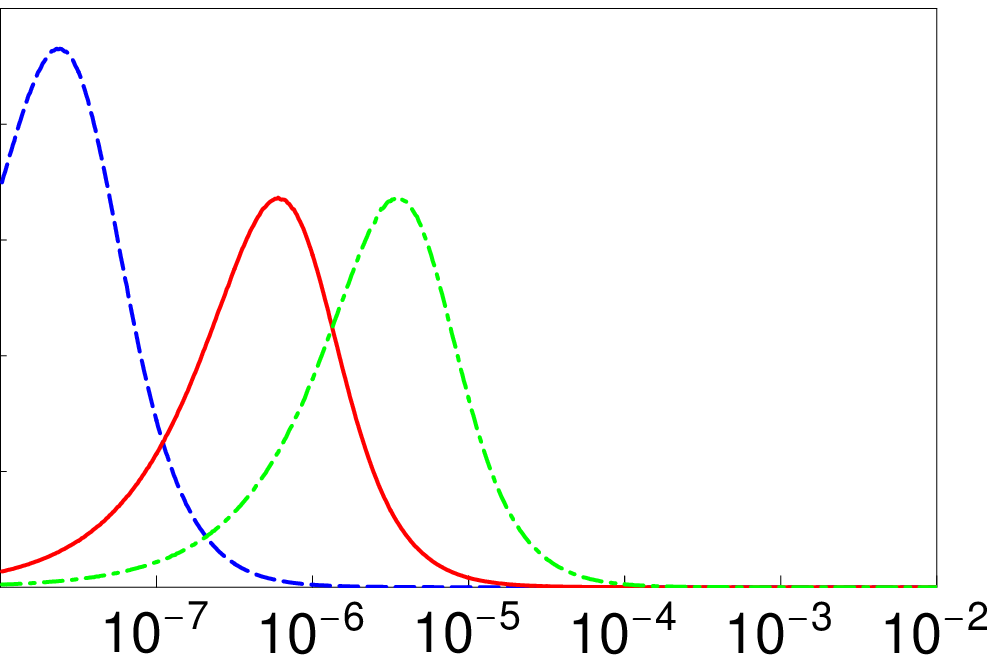,height=6.0cm,width=6.0cm}}}
\end{picture}
\begin{picture}(0,55)
\put(70,0)
{\mbox{\epsfig{file=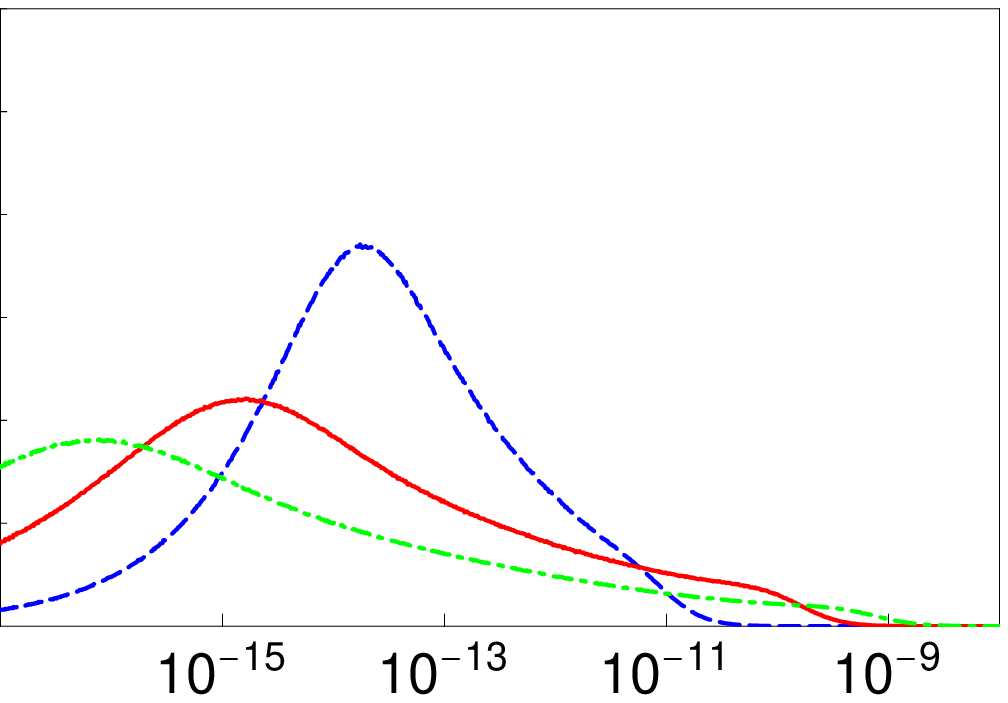,height=6.0cm,width=6.0cm}}}
\end{picture}
{\vskip-10mm\hskip50mm {\Large $\epsilon$} \hskip70mm $Y_B$
\vskip0mm}
\caption[allfig]{Plots of $\epsilon$ (to the 
left) and $Y_B$ (to the right) for the model defined in eq. \ref{caseb}. 
Line styles are as in fig. \ref{fig:modb}.}
        \label{fig:modb2}
\end{figure}

As seen from Fig. \ref{fig:modb} one can adjust the hierarchy 
in the right-handed sector by a rather small change in $\bar{\lambda}$. 
Going from $\bar{\lambda} = \sqrt{\lambda} \simeq 0.47$ to 
$\bar{\lambda} = 0.55$ ($0.60$) changes $M_1$ from 
$ M_1 \sim$ (few) $10^{8}$ GeV to $ M_1 \sim 10^{10}$ 
($10^{11}$) GeV. This way it is possible to achieve larger 
values of $\epsilon$ and a value of $Y_B$ marginally consistent 
with experimental data as shown in Fig. \ref{fig:modb2}. Note, 
however, that the peaks in $Y_B$ for this model are still too 
small and the models survive the leptogenesis test only in the 
tails of the distributions. Although in principle
in models of this kind (case (b))
there is no association of the parameter ${\tilde m}_1^{(b)}$ in 
Eq.\ref{m1b} with a physical neutrino mass, in this case when
we calculate ${\tilde m}_1^{(b)}$ we must first rotate to the
basis in which the right-handed neutrino mass matrix is diagonal.
This will lead to larger values of ${\tilde m}_1^{(b)}\sim 10^{-2}-10^{-1}$ 
eV than would naively be estimated from Eq.\ref{yukb}.

This change in ${\bar \lambda}$ also influences (although only 
rather weakly) the preferred values of $R$. As shown in Fig. \ref{fig:modb} 
this model tends to prefer values of $R$ consistent with the LOW 
solution of the solar neutrino problem. 
As mentioned previously, the LOW solution really only makes sense
within the framework of SRHND because of the large
hierarchies of neutrino masses which would otherwise appear
rather unnatural.

\begin{figure}
\setlength{\unitlength}{1mm}
\begin{picture}(0,55)
\put(0,0)
{\mbox{\epsfig{file=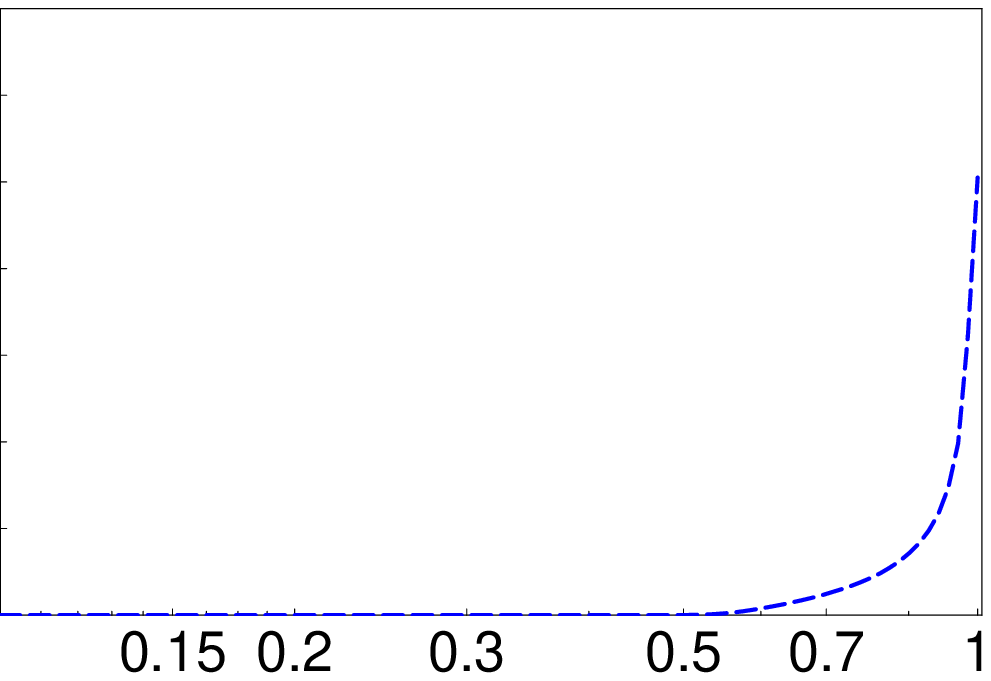,height=6.0cm,width=6.0cm}}}
\end{picture}
\begin{picture}(0,55)
\put(70,0)
{\mbox{\epsfig{file=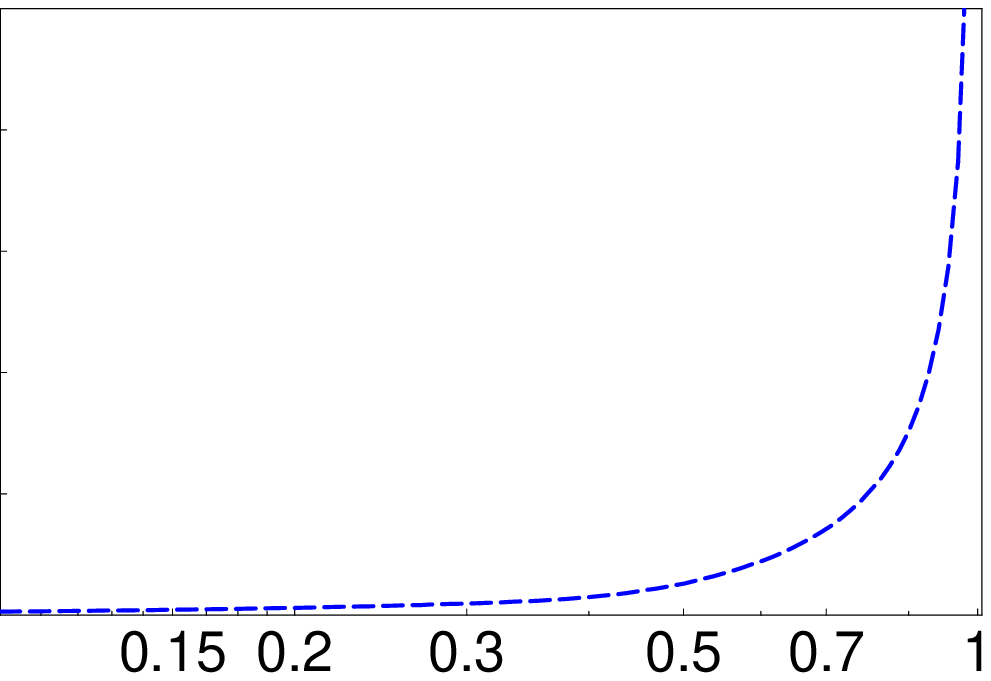,height=6.0cm,width=6.0cm}}}
\end{picture}
{\vskip-10mm\hskip50mm {\Large $s_{Atm}$ \hskip60mm $s_{\odot}$}
\vskip2mm}
\begin{picture}(0,55)
\put(0,0)
{\mbox{\epsfig{file=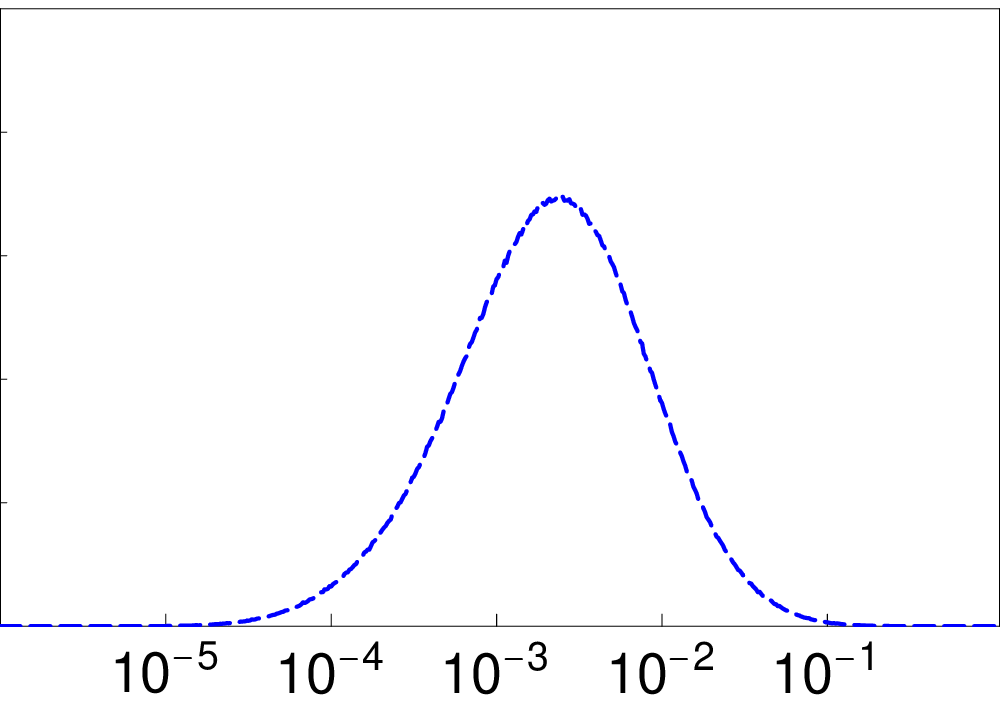,height=6.0cm,width=6.0cm}}}
\end{picture}
\begin{picture}(0,55)
\put(70,0)
{\mbox{\epsfig{file=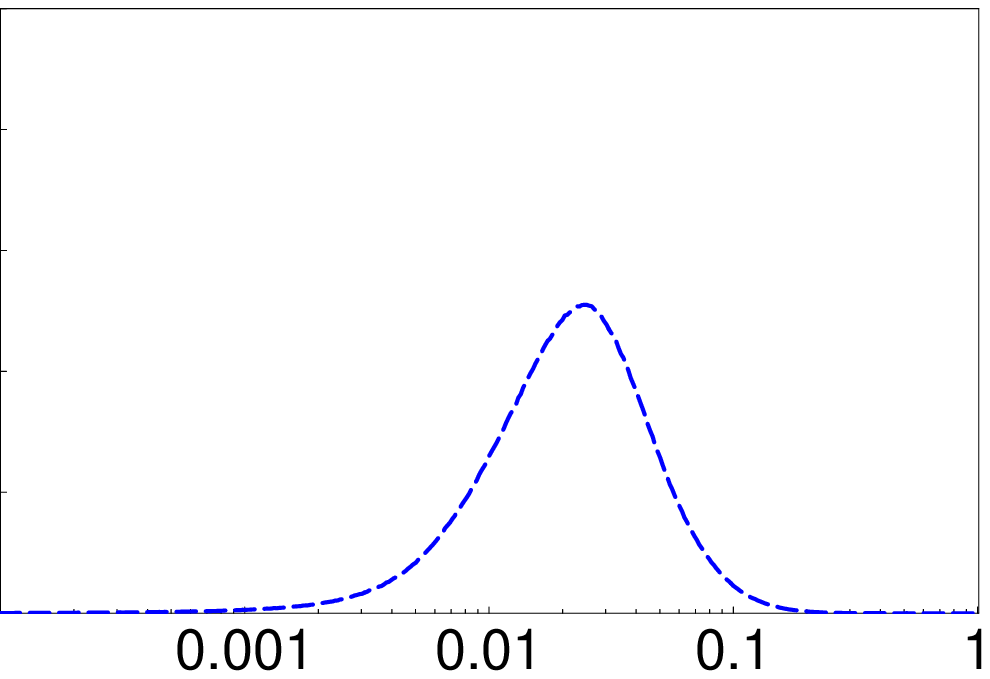,height=6.0cm,width=6.0cm}}}
\end{picture}
{\vskip-10mm\hskip50mm  $R$ \hskip70mm {\Large $s_C$}
\vskip2mm}
\begin{picture}(0,55)
\put(0,0)
{\mbox{\epsfig{file=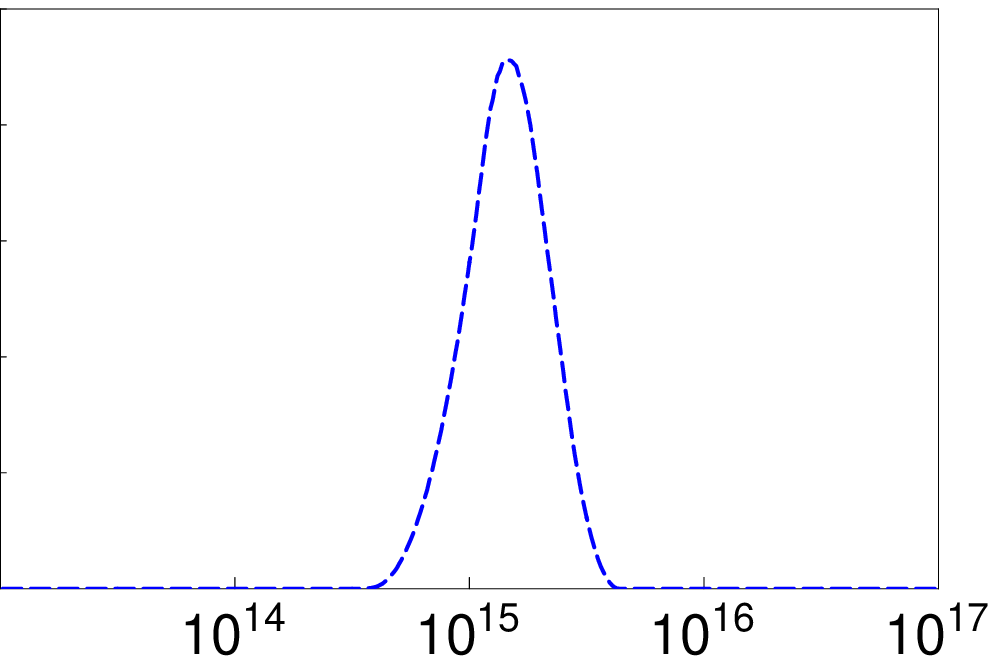,height=6.0cm,width=6.0cm}}}
\end{picture}
\begin{picture}(0,55)
\put(70,0)
{\mbox{\epsfig{file=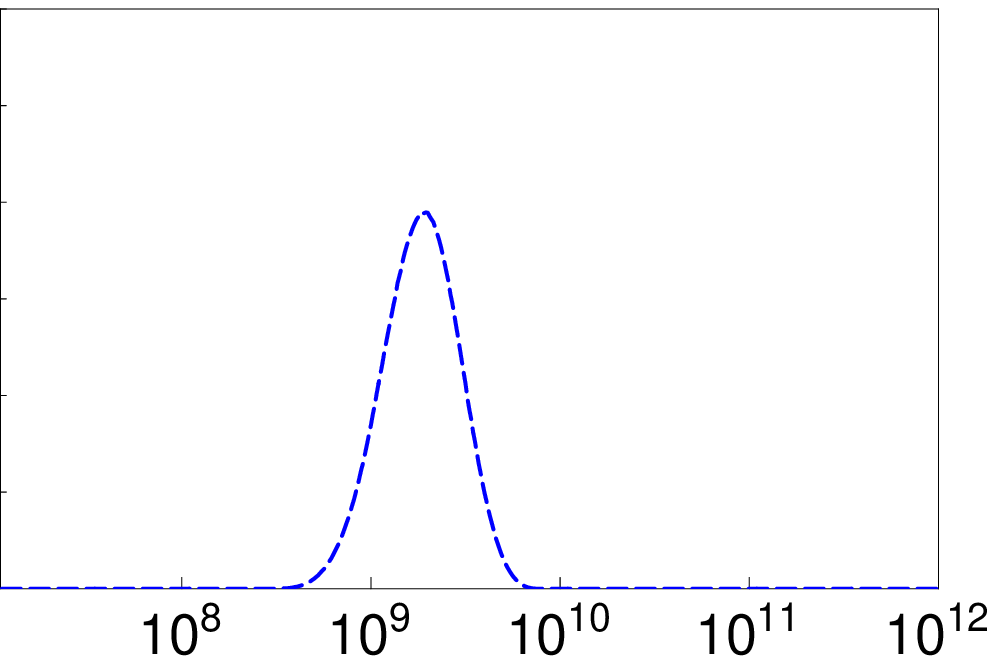,height=6.0cm,width=6.0cm}}}
\end{picture}
{\vskip-10mm\hskip40mm {$\langle \Sigma \rangle$ [GeV]\hskip55mm {$M_1$ [GeV]}}
\vskip2mm}
\caption[allfig]{Plots of (from top left to bottom right: $s_{Atm}$, 
$s_{\odot}$, $R$, $s_C$, $\langle \Sigma \rangle$ and $M_1$ [GeV] 
for the Pati-Salam model discussed in eqs \ref{yuk}-\ref{mrr}. Note 
that this model is consistent with the LA-MSW solution to the solar 
neutrino problem.}
        \label{fig:lam9atm}
\end{figure}

One of the advantages of having the dominant right-handed neutrino
as the heaviest is that leptogenesis may be achieved consistent
with $e \sim 1$, which allows the third (dominant, and heaviest)
right-handed neutrino to be associated with the third family
in unified models. An example of such a model
was recently presented in the framework of a string-inspired
SUSY Pati-Salam (PS) model \cite{steve}.
The model in \cite{steve} will not be repeated here,
but we would emphasise that it was deduced from an analysis
of the quark and lepton masses and mixing angles without any 
consideration of leptogenesis, and therefore we find it somewhat
remarkable that it leads to a baryon asymmetry of the correct
order of magnitude. The model in \cite{steve}
leads to the following structure for the Yukawa and right-handed 
neutrino mass matrix:

\begin{equation}
Y_{\nu}^{PS}  
\sim 
\left( \begin{array}{ccc}
\lambda^{7.5} & \lambda^{3.5} & \lambda^{1.5}    \\
\lambda^{6.5} & \lambda^{3.5} & 1   \\
\lambda^{6.5} & \lambda^{3.5} & 1
\end{array}
\right)
\label{yuk}
\end{equation}
\begin{equation}
M_{RR}^{PS}  
\sim 
\left( \begin{array}{ccc}
\lambda^9 & \lambda^7 & \lambda^5    \\
\lambda^7 & \lambda^5 & \lambda^{3.5}   \\
\lambda^5 & \lambda^{3.5} & 1   
\end{array}
\right)
\label{mrr}
\end{equation}

The effective light Majorana matrix then has contributions from
the third, second and first right-handed neutrinos of:

\beq
m_{LL}^{PS}
\sim 
\left( \begin{array}{ccc}
\lambda^3 & \lambda^{1.5} & \lambda^{1.5}    \\
\lambda^{1.5} & 1 & 1    \\
\lambda^{1.5} & 1 & 1  
\end{array}
\right)
+ {\cal O}
\left( \begin{array}{ccc}
\lambda^2 & \lambda^2 & \lambda^2    \\
\lambda^2 & \lambda^2 & \lambda^2    \\
\lambda^2 & \lambda^2 & \lambda^2   
\end{array}
\right)
+ {\cal O}
\left( \begin{array}{ccc}
\lambda^6 & \lambda^5 & \lambda^5    \\
\lambda^5 & \lambda^{4} & \lambda^{4}    \\
\lambda^5 & \lambda^{4} & \lambda^{4}   
\end{array}
\right).
\label{SRHND5}
\eeq

\begin{figure}
\setlength{\unitlength}{1mm}
\begin{picture}(0,55)
\put(0,-10)
{\mbox{\epsfig{file=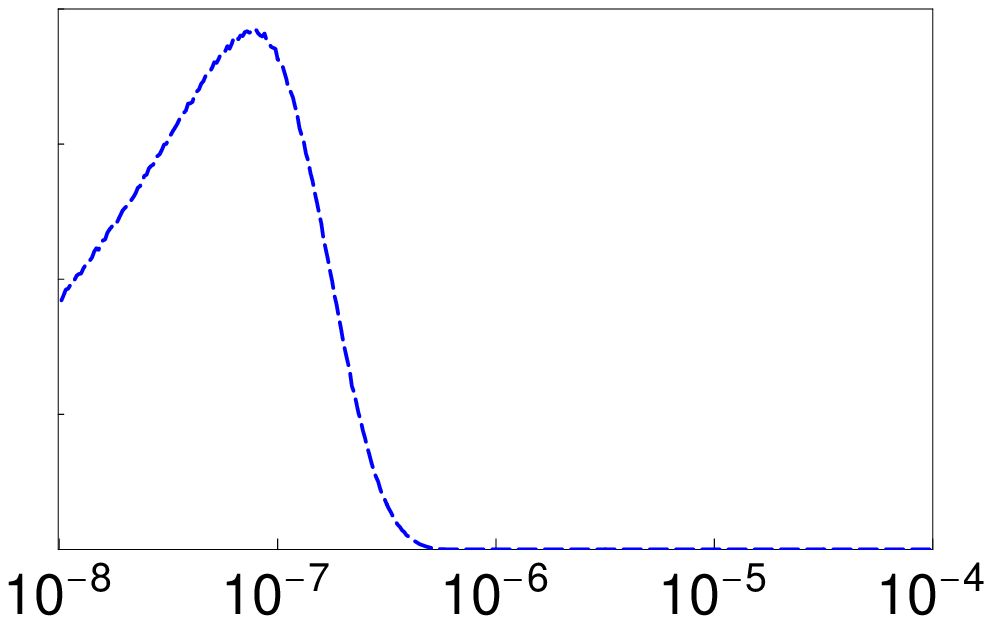,height=6.0cm,width=6.0cm}}}
\end{picture}
\begin{picture}(0,55)
\put(70,-10)
{\mbox{\epsfig{file=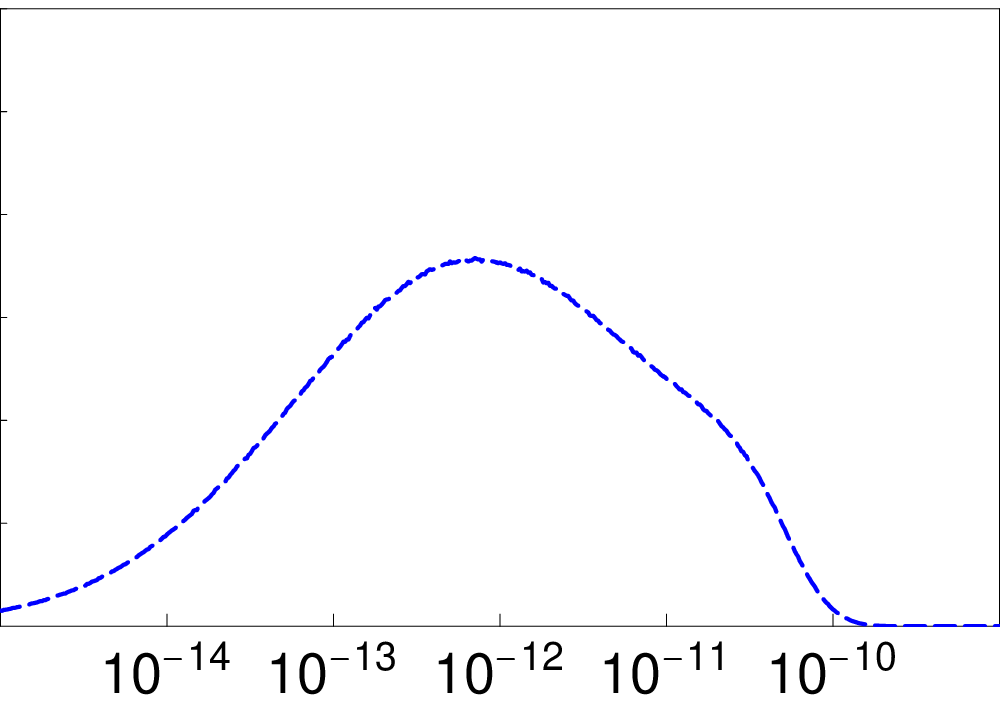,height=6.0cm,width=6.0cm}}}
\end{picture}
{\vskip-1mm\hskip50mm {\Large $\epsilon$} \hskip70mm $Y_B$
\vskip2mm}
\caption[allfig]{Plots of $\epsilon$ (to the 
left) and $Y_B$ (to the right) for the Pati-Salam model of \cite{steve}.}
        \label{fig:lam9eps}
\end{figure}

From the analytic estimates in 3.1 we expect this model to be consistent
with the LMA MSW solution. This is explicitly demonstrated in 
Fig. \ref{fig:lam9atm}. From the analytic estimates in 3.2
we also expect this model to give successful leptogenesis,
and this is demonstrated in Fig. \ref{fig:lam9eps}.
From the matrices given above and from the analytical estimates of 
eq. \ref{b} one expects that $\epsilon_1 \sim 3/(16\pi) \times \lambda^9 
\sim 7 \times 10^{-8}$, which within a factor of $\sim 2$ or so agrees with 
the numerical calculation of $\epsilon_1$ in
fig. \ref{fig:lam9eps}. Note, that the 
resulting values of $Y_B$ in fig. \ref{fig:lam9eps} 
are also consistent with the arguments in 3.2.
In particular in such case (b) models where the dominant right-handed
neutrino is the heaviest it is easier to avoid the gravitino constraint
\cite{gravitino}, although the values of $M_1$ are still a bit 
on the high side, as seen in fig. \ref{fig:lam9atm}.
As in the previous model, in the diagonal right-handed neutrino
basis we obtain a larger value of ${\tilde m}_1^{(b)}\sim 10^{-2}$ eV
than would naively be estimated from Eq.\ref{yuk}.

Note that supersymmetric models of case (b) have the feature that there
is an order unity Yukawa coupling in the 23 position of the Yukawa
matrix which leads to a large off-diagonal entry in the slepton
mass matrix. This leads to the striking signature of the
lepton flavour violating (LFV) process
$\tau \rightarrow \mu \gamma$ close to the 
experimental upper limit, as first pointed out in \cite{blazek}.
In general LFV constraints 
provide an additional window into the see-saw matrices
in supersymmetric models \cite{blazek,LFV}.

\section{Summary and Conclusions}

This paper represents the first study of leptogenesis based on 
hierarchical models of neutrino masses in which SRHND is used to 
generate the neutrino mass hierarchy. Such models have been shown to 
accomodate the presently favoured large solar angle solutions such 
as LMA and LOW \cite{SRHND2}, and in the case of the LOW solution 
where the neutrino mass hierarchy is large it would seem that SRHND 
is almost inevitable.
So we would argue that, far from this analysis being restricted
to a particular small class of models, it is in fact
quite generally applicable to large classes of models in which
the neutrino mass hierarchy is generated in a natural way without
any fine-tuning. Therefore the above results should be regarded
as being quite generally applicable to see-saw models 
containing a neutrino mass hierarchy.

In presenting our analytic and numerical results
we make a clear distinction between the theoretically clean
asymmetry parameter $\epsilon_1$ and the baryon asymmetry $Y_B$,
for which we present and use a fit based on a Boltzmann analysis.
We have presented analytic expressions for both the MNS
parameters, extending the previously presented analytic results
\cite{SRHND2} to the complex domain, and for leptogenesis
asymmetry parameter $\epsilon_1$ in the cases where the dominant
right-handed neutrino is either the heaviest or the lightest.
We have compared the analytic estimates to full numerical
results for models based on $U(1)$ family symmetry,
and have performed
a numerical scan over the unknown coefficients, and have seen that
the peaks of the distributions in $\epsilon_1$ 
are in good agreement with the analytic results.
Using the analytic and numerical approaches we then 
discussed leptogenesis decoupling and leptogenesis discrimination.

We have shown that quite generally there is a decoupling between
the low energy neutrino observables and the leptogenesis predictions
for $\epsilon_1$.
Thus leptogenesis has nothing to tell us about which
solar solution we would expect, and for example the
LMA and the LOW solutions are equally acceptable,
as indeed would have been the SMA solution were it not disfavoured
by SNO and Super-Kamiokande. Furthermore the leptogenesis
phase is independent of the measurable MNS phase, although the
analytic estimates make it clear that since the two phases
originate from the same Yukawa matrix, and even in some
cases involve the phases of the same Yukawa couplings,
the general expectation is that, barring cancellations,
both sorts of phases should be of roughly the same order
of magnitude. 

In going from $\epsilon_1$ to $Y_B$ one needs to 
make some assumptions concerning the cosmological
history of the universe. In this paper we have assumed a
standard hot big bang universe, which is equivalent to 
assuming a very high reheat temperature after inflation
which is larger than the  right-handed neutrino masses.
Within this standard cosmology the right-handed neutrinos
are produced via their couplings to the thermal bath,
yet they are required to decay out-of-equilibrium,
leading to a rather narrow range of couplings ${\tilde m}_1$
of the lightest right-handed neutrino
consistent with successful leptogenesis.
For the calculation of $Y_B$ 
a correct treatment of the Boltzman equations 
describing the number evolution of the heavy right-handed neutrinos 
in the early universe is essential \cite{FitExact}. We therefore have 
devised an empirical fit formula and compared it to the exact results 
\cite{FitExact} as well as to the simpler approximation \cite{kolb,yasutaka}. 
Although for a small range in ${\tilde m}_1$ and small values of the 
lightest right-handed neutrino mass the simple approximation 
\cite{kolb,yasutaka} agrees reasonably with the exact result \cite{FitExact}, 
for most parts of the parameter space it fails badly. Only if one 
takes into account the suppresion of the dilution factor $d$ for 
larger values of ${\tilde m}_1$ and $M_1$, either by solving the 
Boltzman equations numerically or by the use of our approximate fit 
formula, does one find reliable results. Without taking this effect 
into account we would have wrongly concluded that $Y_B$ can get 
as large as $Y_B \sim 10^{-5}$ in some models, whereas with our more 
refined treatment we find that $Y_B$ is always $Y_B \le 10^{-14}$, 
if the dominant right-handed neutrino is the lightest. 

Based on the above analysis of $Y_B$
we have shown that leptogenesis excludes a large class of models where
the dominant right-handed neutrino is the lightest one.
The power of leptogenesis to resolve ambiguities 
between models which would otherwise lead to the same neutrino observables
provides a welcome constraint on high energy theories. 
We have shown that models where the dominant right-handed neutrino is the 
heaviest are marginally consistent with the gravitino constraint
and have studied an explicit example of a unified model of this type.
We find it encouraging that a model which was written down to describe
the fermion mass spectrum \cite{steve}, including the neutrino 
spectrum and the LMA MSW solution, should be precisely of this
kind and gives successful leptogenesis, subject to the uncertainties
of our estimates discussed in section 2.3.

Finally we should emphasise that our conclusions are
based on the assumed cosmological history being the standard
hot big bang with a high reheat temperature.
One plausible alternative is to suppose that the reheat temperature
is below $10^9$ GeV, but that heavier right-handed (s)neutrinos 
can be produced in sufficient numbers by preheating at the end of
inflation \cite{preheating}. 
The preheating must efficiently produce right-handed
(s)neutrinos without over-producing gravitinos, and this will depend
on the precise details of the inflation model. A model of leptogenesis
with a low reheat temperature, based on preheating of heavy
right-handed sneutrinos, which does not suffer from the gravitino problem
has been recently studied in detail in \cite{ma_king}.
The same Pati-Salam model has also been studied in this context
\cite{ma_king} and interestingly the results for $Y_B$
are also consistent, within the large uncertainty, 
with the estimates given here,
based on an entirely different
cosmological history of the universe.

\bigskip
\begin{center}
{\bf Acknowledgements}
\end{center}
{\small We thank M.Pl\"umacher for various discussion on the numerical 
solution to the Boltzman equations. S.K. is grateful to PPARC for the
support of a Senior Fellowship. }

\end{document}